\documentclass[fleqn,usenatbib]{mnras}

\usepackage{newtxtext,newtxmath}

\usepackage[T1]{fontenc}

\DeclareRobustCommand{\VAN}[3]{#2}
\let\VANthebibliography\thebibliography
\def\thebibliography{\DeclareRobustCommand{\VAN}[3]{##3}\VANthebibliography}

\usepackage{graphicx}	% Including figure files
\usepackage{float}
\usepackage{amsmath}	% Advanced maths commands
\usepackage{gensymb}
\usepackage{flafter}
\usepackage[normalem]{ulem}
\usepackage[section]{placeins}
\usepackage{subcaption}

\title[Diurnal temperature variation of Mars]{Diurnal variation of the surface temperature of Mars with the Emirates Mars Mission: A comparison with Curiosity and Perseverance rover measurements}

\author[Atri et al.]{Dimitra Atri\thanks{Email: atri@nyu.edu}, Nour Abdelmoneim, Dattaraj B. Dhuri, Mathilde Simoni
\\
Center for Space Science, New York University Abu Dhabi, PO Box 129188, Saadiyat Island, Abu Dhabi, UAE}

\date{Accepted XXX. Received YYY; in original form ZZZ}

\pubyear{2022}

\begin{document}
\label{firstpage}
\pagerange{\pageref{firstpage}--\pageref{lastpage}}
\maketitle

%Target journal MNRAS  
\begin{abstract}
For the first time, the Emirates Mars Infrared Spectrometer (EMIRS) instrument on board the Emirates Mars Mission (EMM) ``Hope", is providing us with the temperature measurements of Mars at all local times covering most of the planet. As a result, it is now possible to compare surface temperature measurements made from orbit with those from the surface by rovers during the same time period. We use data of diurnal temperature variation from the Rover Environmental Monitoring Station (REMS) suite on board the Mars Science Laboratory (MSL) ``Curiosity" rover, and the Mars Environmental Dynamics Analyzer (MEDA) suite on board the Mars 2020 ``Perseverance" rover, between June and August 2021 and compare them with EMIRS observations and estimates of the Mars Climate Database (MCD) model. We show that although the overall trend of temperature variation is in excellent agreement across missions, EMIRS measurements are systematically lower at night compared to Mars 2020. The lower spatial resolution of EMIRS compared to the rovers and consequently lower average thermal inertia of the observed regions in this particular case primarily contributed to this discrepancy, among other factors. We discuss the implications of these results in improving our understanding of the Martian climate which would lead to better modeling of local weather prediction, useful for future robotic and crewed missions. 
\end{abstract}


%%%%%%%%%%%%%%%%%%%%%%%%%%%%%%%%%%%%%%%%%%%%%%%%%%

%%%%%%%%%%%%%%%%% BODY OF PAPER %%%%%%%%%%%%%%%%%%

\section{Introduction}
The transformation of Mars from a warm, wet planet with potentially habitable conditions to a cold and dry planet that we see today is an active area of research by space agencies around the globe \citep{vago2015esa, zurbuchen2017mars, bhardwaj2014indian, amiri2022emirates}. It is not only important to investigate this in order to understand Mars, but for enabling us to better understand the evolution of earth and other planets in the Solar system and beyond \citep{sagan1972earth, jakosky2021atmospheric}. A number of atmospheric and surface processes occurring on Mars at present provide us with clues to this long-term transformation of the planet. Studying the diurnal and seasonal variability of the surface temperature of Mars is of prime importance because of its key role in governing a number of surface, shallow subsurface and lower atmospheric processes of the planet and its influence on weather, climate and the eventual escape of volatiles from the atmosphere \citep{petrosyan2011martian}. 

Surface temperature measurements have been made in the past from orbiters, such as the Thermal Emission Spectrometer (TES) instrument on board the Mars Global Surveyor (MGS) \citep{christensen2001mars}, the Planetary Fourier Spectrometer (PFS) instrument \citep{formisano2005planetary} on board Mars Express (MEX), the Thermal InfraRed channel in honor of professor Vassilii Ivanovich Moroz (TIRVIM), a part of the Atmospheric Chemistry Suite (ACS) onboard the ExoMars Trace Gas Orbiter (TGO) \citep{korablev2018atmospheric}, Mars Reconnaissance Orbiter/Compact Reconnaissance Imaging Spectrometer for Mars(MRO/CRISM) \citep{he2022surface}, Mars Odyssey/Thermal Emission Imaging System (MO/THEMIS) \citep{edwards2018thermophysical}. Such measurements have also been made by landers and rovers, such as the Viking landers 1 and 2 \citep{hess1977meteorological}, Mars Exploration Rovers (MER), Spirit and Opportunity \citep{spanovich2006surface}\citep{mason2021temperature}, Rover Environmental Monitoring Station (REMS) on board the Mars Science Laboratory (MSL) ``Curiosity" \citep{gomez2012rems}\citep{martinez2021surface}, Mars Environmental Dynamics Analyzer (MEDA) on board the Mars 2020 ``Perseverance" rover \citep{rodriguez2021mars}\citep{martinez2022thermal}, and the Heat flow and Physical Properties Package (HP3) for the InSight mission \citep{spohn2018heat}\citep{piqueux2021soil}. Depending on the orbit, satellite measurements have the advantage of geographical coverage, they suffer from a lack of local time coverage. On the other hand, rovers provide measurements at all local times, but the measurements are confined geographically. The Emirates Mars Mission (EMM) or ``Hope" aims to fill-in this gap in observations \citep{almatroushi2021emirates}.  

One of the main goals of EMM is to provide a global view of Mars on diurnal and seasonal timescales. EMM's unique orbit enables it to observe the spatial distribution of surface and near-surface temperature for most of the planet at all local times \citep{amiri2022emirates, almatroushi2021emirates, edwards2021emirates}. It also enables measurements of diurnal temperature variations over full Martian day with near-complete coverage of the entire planet.  Since we are able to observe diurnal temperature variations at most locations on Mars, it opens an opportunity to directly compare observations from the orbit and from the surface of Mars. 

In this letter, we report diurnal temperature variation for the most of the planet measured by EMM from orbit and compare it with surface measurements made by MSL \citep{gomez2012rems} and Mars 2020 \citep{rodriguez2021mars} rovers over the same time period. We describe a number of factors which could lead to these differences in measurements and propose solutions. We also compare these measurements with the Mars Climate Database (MCD) \citep{millour2018mars} and discuss implications for improved modeling of Martian weather and climate.
%Diurnal temperature variations are important to study because of their role in governing a number of surface and lower atmospheric processes and would lead to better constrain various climate models and also upper atmospheric processes such as atmospheric escape, which are closely tied to surface and near-surface processes.
\section{Spacecraft measurements}
\subsection{Orbiter: EMIRS/EMM}
The Emirates Mars Infrared Spectrometer (EMIRS) instrument \citep{edwards2021emirates} on board EMM is a Fourier Transform Infra Red (FTIR) spectrometer observing in 6-100 $\mu$m wavelength range. It is designed to provide measurements of the lower atmosphere and surface of Mars by conducting thermal infrared observations of the disk. It provides diurnal measurements over the entire globe on sub-seasonal timescales. EMM's orbital period is $\sim$ 55 hours with a periapsis of 20,000 km and an apoapsis of 43,000 km. It takes $\sim$ 20 observations per orbit or $\sim$ 60 per week with one observation taking 29 minutes at periapsis and 11 minutes at apoapsis. It takes 4 orbits or less than 10 days to obtain the global coverage at all local times. The resolution ranges between 100 to 300 km/pixel (diameter) with up to 70$\degree$ emission angle. Level 2 data products (L2) contain calibrated radiance and brightness temperature. Level 3 products (L3), used in this study, are derived from L2 and have corrected surface temperatures among other quantities. Surface temperatures (L3) are obtained using a three-step process \citep{edwards2021emirates}: first the atmospheric temperatures are retrieved from the 15 $\mu$m CO$_2$ band, then dust optical depth and water-ice aerosol optical depths are fit, and finally the water vapor column depth is fit. This retrieval process is iterated until convergence, and quality control metrics are used to ensure the quality of final retrieved products. A comparison of L2 and L3 measurements is shown in the appendix figure A1.

%Description of data analysis
The orbiter was inserted in the Martian orbit on February 9, 2021 (MY 36, L$_{s}$ = 0.6) providing 9 sols of EMIRS observations for the month from the insertion orbit. The orbiter was in transition orbit for the next couple of months, starting the science phase of the mission from May 23, 2021 (L$_{s}$ = 48.7). %After preprocessing the data and excluding observations with boresights that did not intersect with the target (bore flag = 0), there were 5 days of data in the month of May (4410 measurements), 19 days in June (20470 measurements), 31 days in July (34053 measurements), and 25 days in August (16203 measurements). \footnote{EMM SDC data is organized in earth months}.  

\begin{figure}
\centering
	\includegraphics[width=6.25cm]{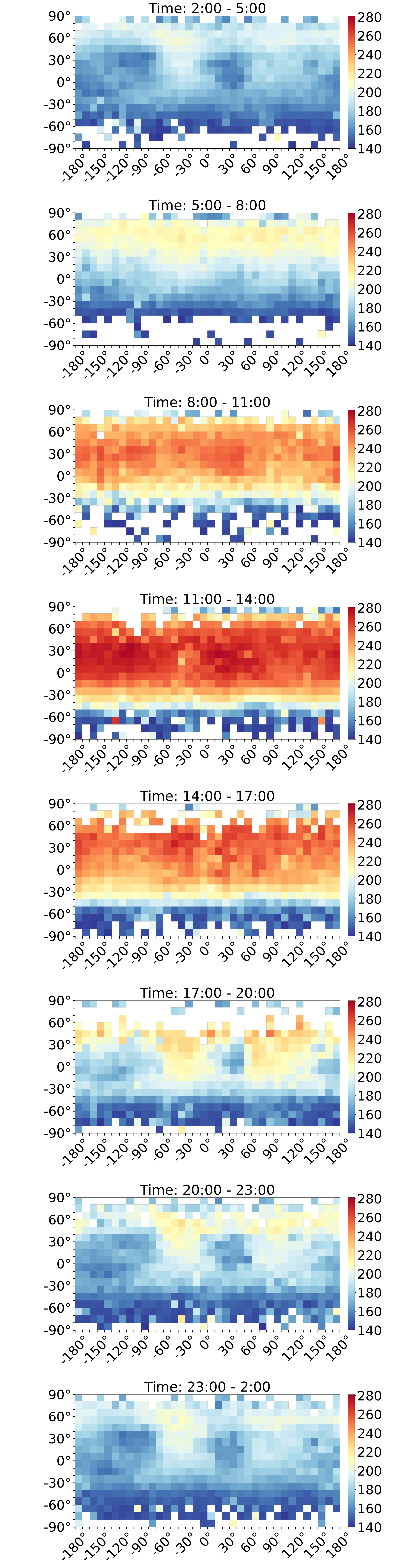}
    \caption{Global surface temperature map based on EMIRS/EMM observations of 30 sols L$_{s}$ = 65.8-78.9 (July 2021). }
    \label{fig:surface_map_july}
\end{figure}

Figure \ref{fig:surface_map_july} shows the diurnal surface temperature variation over 30 sols for L$_{s}$ = 65.8-78.9 (July 2021). The horizontal and vertical axes show latitude and longitude in degrees respectively. The surface temperature is color-coded in units of Kelvin and the time shown is the Local True Solar Time (LTST). We constructed these maps by dividing the planet into $10\degree \times 10\degree$ latitude/longitude bins. Appendix figure B1 shows the number of samples in each spatial bin. We then obtained a temperature value for each bin by averaging the surface temperature measurements during each of the 3-hour time intervals. In both the polar regions, data above 80$\degree$ is scarce. For the northern polar region, the data is very limited between 14:00 - 20:00 hours and for the southern polar region, data is very limited between 02:00 - 14:00 hours. One can easily notice the large-scale variation in surface temperature across the planet in a given time slot, and also global changes in temperature over the course of the sol. Temperatures in the northern hemisphere are systematically higher corresponding to the summer season. As expected, the daily temperature variability is most noticeable in the northern mid-latitudes. Global temperature maps for other months are available in the appendix figures B2-B4. 
%Figures \ref{fig:surface_hours_msl_july}-\ref{fig:surface_hours_percy_aug} show that hourly surface temperature measurements from EMIRS are in good agreement with REMS and MEDA.  

\begin{figure}
\centering
	\includegraphics[width=8.5cm]{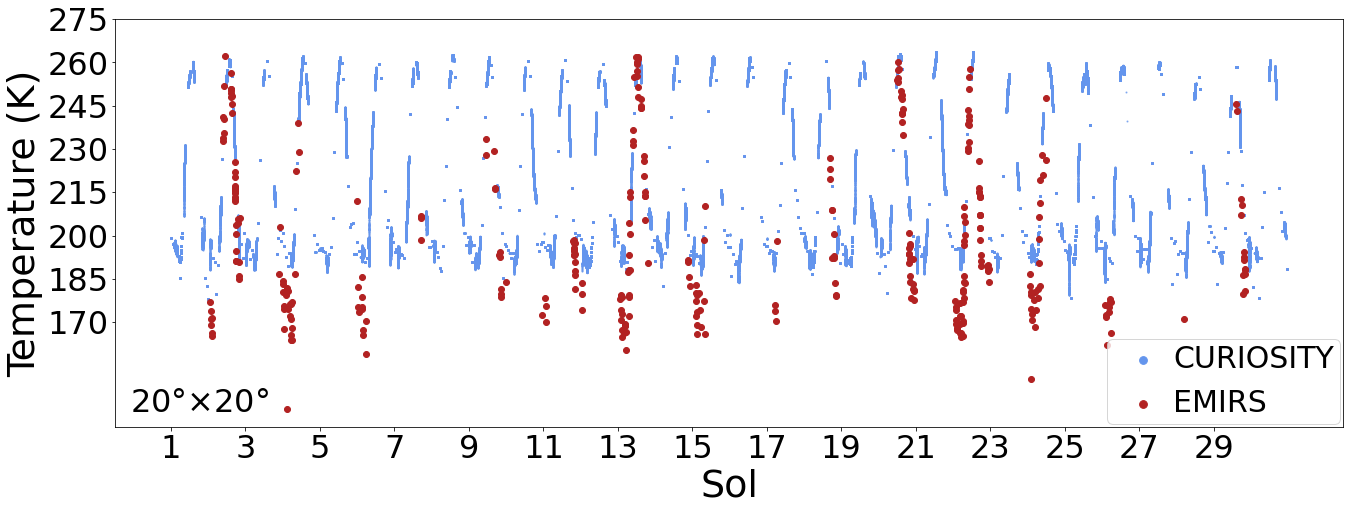}\\
	\includegraphics[width=8.5cm]{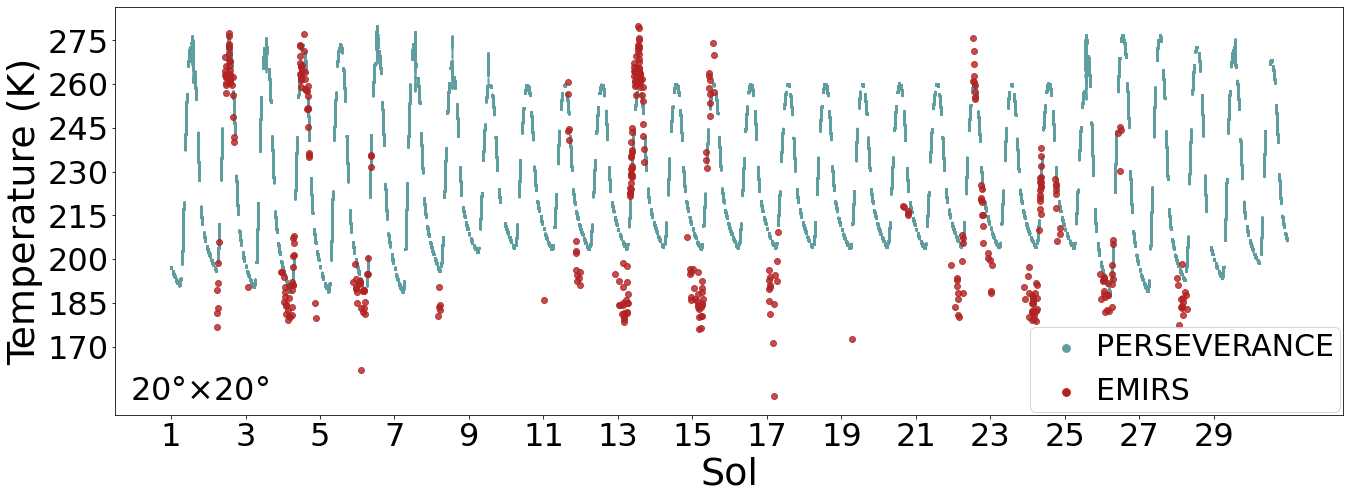}\\
	\includegraphics[width=8.5cm]{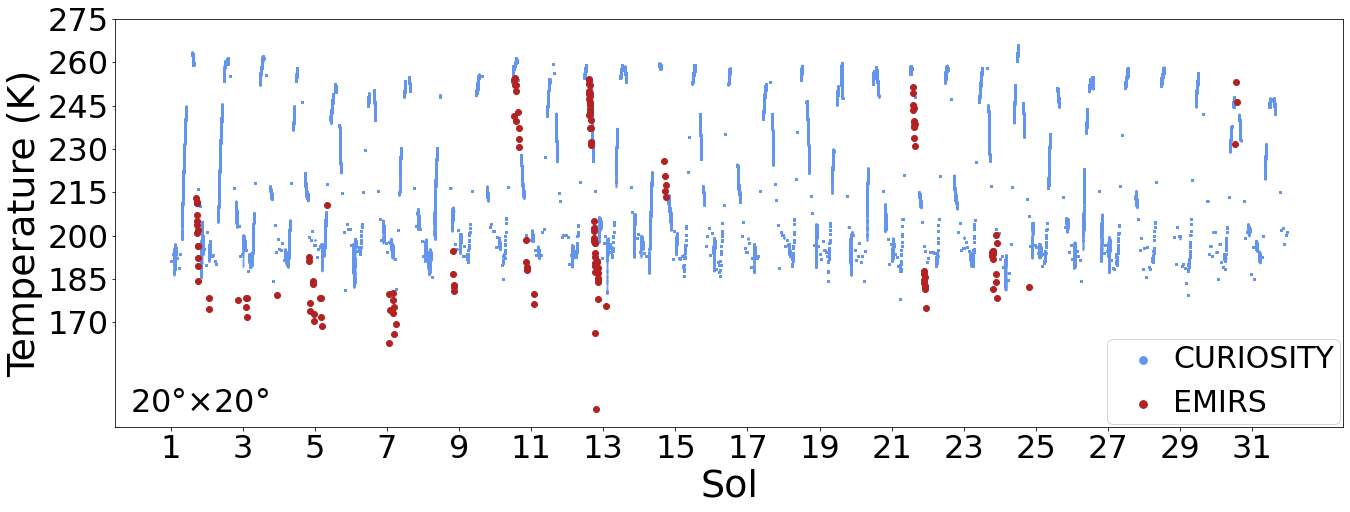}\\
	\includegraphics[width=8.5cm]{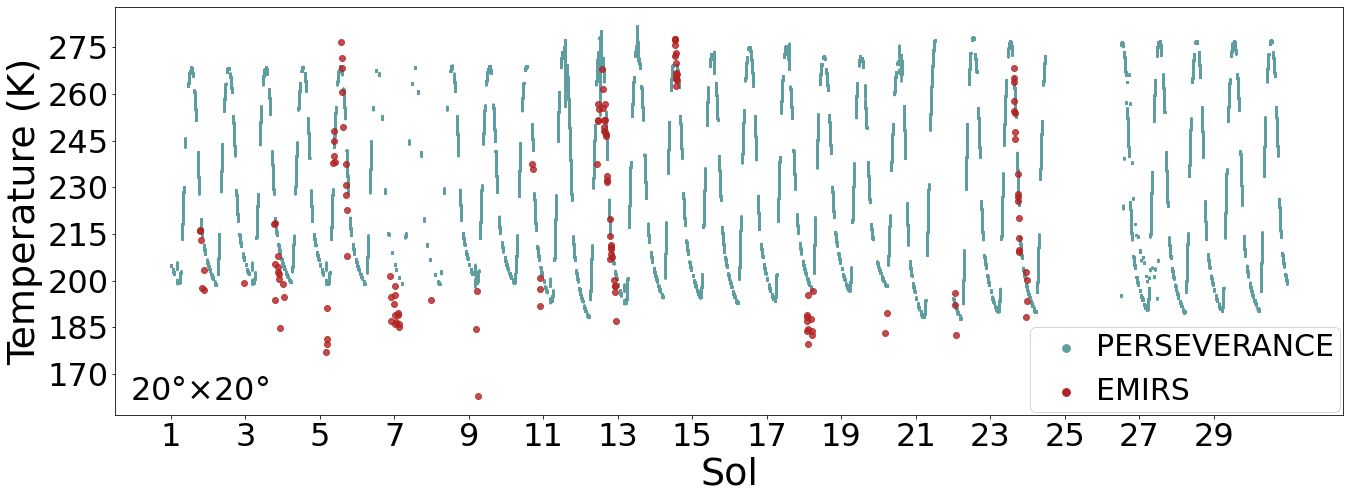}
    \caption{Comparison of hourly surface temperatures between EMIRS, REMS/MSL and MEDA/Mars 2020 over $\sim$~30~sols for L$_{s}$ = 65.8 -78.9 (July 2021, {\color{red} sols 3164-3194 (REMS/MSL) and sols 128-158 (MEDA/Mars 2020)}, {\it top two panels}) and for L$_{s}$ = 79.3-92.5 (August 2021, {\color{red} sols 3194-3224 (REMS/MSL) and sols 158-188 (MEDA/Mars 2020)}, {\it bottom two panels}). Each point corresponds to a single observation and gaps indicate missing observations.}
    \label{fig:surface_hours_comparisons}
\end{figure}

\subsection{Surface rover: REMS/MSL}             
The Rover Environmental Monitoring Station (REMS) \citep{sebastian2010rover} instrument on board MSL provides meteorological measurements at the rover's location in Gale crater on diurnal and seasonal scales \citep{gomez2012rems}. Gale crater, the operating site of MSL, is located in the Aeolis Quadrangle at the edge of the dichotomy boundary. It has a diameter of 154 km, is 4 km below datum at the landing site and features Mount Sharp, which is 1.5 km above datum. Ground and air temperature measurements are available over periods overlapping with EMM observations. Measurements are taken by two orthogonally placed booms -- Boom 1 hosts the Air Temperature Sensor (ATS) and the Ground Temperature Sensor (GTS), and Boom 2 hosts another ATS. GTS, which is mounted on Boom 1, measures the IR brightness temperature of the surface with three thermopiles. REMS measures the ground brightness temperature in the 150-300K range with an accuracy of 1 K and a resolution of 2 K \citep{martinez2017modern}. The air temperature is also measured by both the booms with the same accuracy and resolution.

The initial data was obtained from MODRDR (Models RDR) files containing the REMS instrument's highest level data product for every sol. The  ground brightness temperature as a function of LTST were extracted from sols 3018 to 3224, a total of 207 sols for MY 36, L$_{s}$ = 1.1 - 92.5. An initial cleaning of the dataset revealed an average 1959 null values per sol, ranging from 845 to 10246 rows. This resulted in an average of 23180 readings taken at a rate of 1 Hz for each sol. The data points were first grouped by month (to be consistent with EMM SDC) and subsequently by hour intervals during the day. For each hour interval, the mean, minimum and maximum values as well as the standard error were calculated and plotted on the following graphs.

\subsection{Surface rover: MEDA/Mars 2020}
The Mars Environmental Dynamics Analyzer (MEDA) suit of sensors on board the Mars 2020 rover provides continuous meteorological data from the Jezero crater, including surface temperature measurements \citep{rodriguez2021mars}. Jezero crater has a 49 km diameter and is located in the Isidis basin (1500 km) on the dichotomy boundary. The near-surface atmospheric circulation is influenced by the local and regional topography \citep{newman2021multi, pla2020meteorological, newman2022dynamic}. The Thermal InfraRed Sensor (TIRS) \citep{sebastian2021thermal} is a part of MEDA and is an Infrared (IR) radiometer measuring the brightness temperature of the surface. In particular, IR5 provides the ground temperature, measuring in the 8-14 $\mu$m band, with a dynamic range of 173-293 K, an accuracy of $\pm$ 0.75 K, and a resolution of 0.08 K.

 The ground brightness temperature as a function of LTST were extracted from TIRS between sols 15 to 188, corresponding to MY 36, L$_{s}$ = 13-92. While data was available starting February 18 2021, TIRS was not deployed for the first few sols, therefore, we started our analysis from March 6. We found that data points were missing for 7 sols: sols 23, 24, 25, 26, and 27 in March, sol 97 in May and sol 183 in August, leading to a total of 167 sols included in the analysis. Time intervals corresponding to sol 168 (14:00-16:30), sol 177 (14:00-15:00) and sols 159, 161, 163, 167, 169, and 171 (03:00-04-00) were then removed from the analysis due to calibration activities which resulted in values that are not representative \cite{martinezp}. This initial cleaning revealed an average of 45518 data points per sol, ranging from 4207 to 59135 rows and including 0 null values. The data points were first grouped by month (to keep it consistent with EMM SDC) and subsequently by hour intervals during the day. For each hour interval, the mean, minimum and maximum values as well as the standard error were calculated and plotted on the following graphs.

\subsection{The Mars Climate Database MCD}
The Mars Climate Database (MCD) provides averaged data on the main meteorological variables such as pressure, temperature, atmospheric density and winds \cite{millour2018mars}. It is a derived from Laboratoire de Météorologie Dynamique LMD-GCM \cite{wolfgang2015rocky}, a Global Climate Model (GCM) of Mars, which includes a water cycle model, a chemistry model, a thermosphere model and an ionospheric model. It provides day-to-day variability of main meteorological variables and is widely used in the Mars community. In addition to the output from GCM, it provides high-resolution environmental data, year-to-year variability, dust content variations, seasonal and diurnal cycles of key meteorological variables. We obtain the diurnal variability of surface temperatures from MCD at the sites of the Curiosity and Perseverance rovers at Gale and Jezero crater respectively, in order to compare with the temperature observations from EMIRS. Appendix figure C1 shows locations of the Curiosity and Perseverance rovers.
\begin{figure}
	\includegraphics[width=\columnwidth]{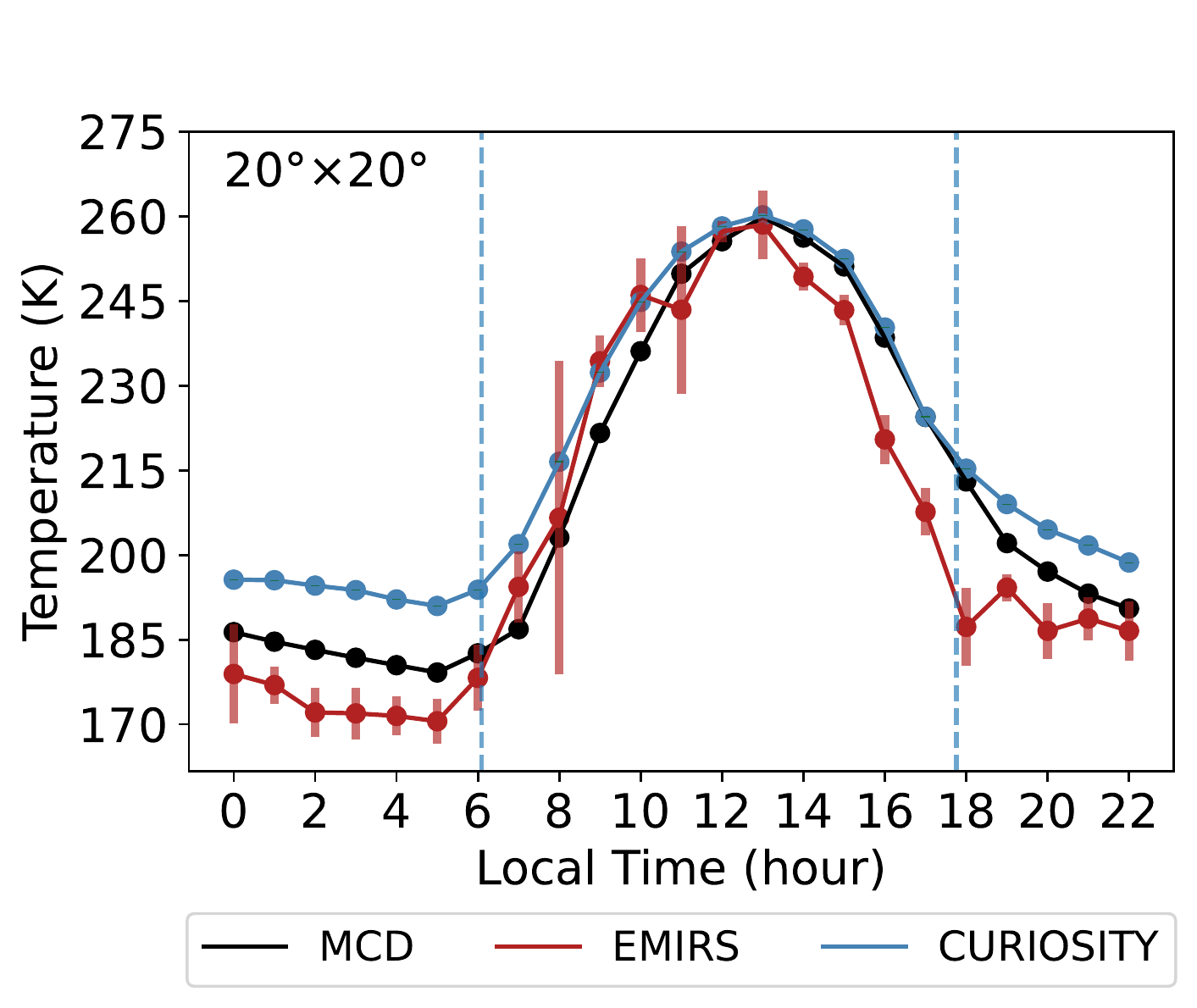}\\
 	\includegraphics[width=\columnwidth]{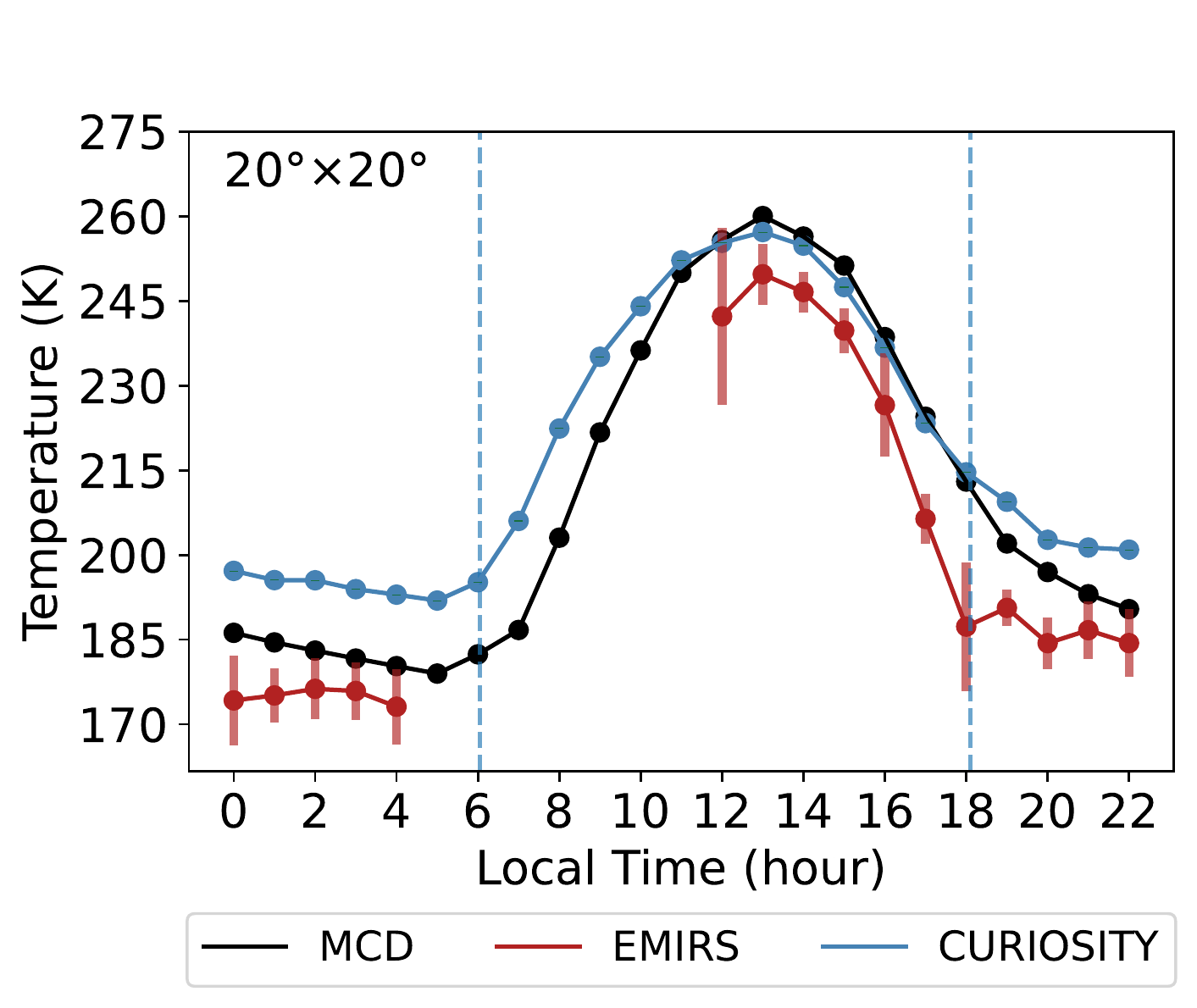}
    \caption{A comparison of average diurnal temperature variation measured by EMIRS/EMM and REMS/MSL with the model prediction from MCD. {\it Top panel}: Over $\sim$~30~sols for L$_{s}$ = 65.8-78.9 (July 2021, {\color{red} sols 3164-3194}). {\it Bottom panel}: Over $\sim$~30~sols for L$_{s}$ = 79.3-92.5 (August 2021, {\color{red} sols 3194-3224}). The error bars indicate 3-$\sigma$ standard errors. Local dawn and dusk are marked by vertical lines. }
    \label{fig:emirs_curiosity}
\end{figure}

\section{Comparison of Measurements}
\subsection{EMM, MSL and MCD}

%MSL figures
% \begin{figure}
% 	\includegraphics[width=\columnwidth]{SurfJuly131_EMIRS_msl_MCD_3stderr-2.pdf}
%     \caption{A comparison of average diurnal temperature variation over 31 sols measured by EMIRS/EMM and REMS/MSL from L$_{s}$ = 65.8 (July 2021) with the model prediction from MCD. The error bars indicate 3-$\sigma$ standard errors. Local dawn and dusk are marked by vertical lines. }
%     \label{fig:emirs_curiosity_july}
% \end{figure}

% \begin{figure}
% 	\includegraphics[width=\columnwidth]{SurfAugust131_EMIRS_msl_MCD_3stderr.pdf}
%     \caption{A comparison of average diurnal temperature variation over 31 sols measured by EMIRS/EMM and REMS/MSL from L$_{s}$ = 79.3 (August 2021) with the model prediction from MCD. The error bars indicate 3-$\sigma$ standard errors. Local dawn and dusk are marked by vertical lines. }
%     \label{fig:emirs_curiosity_august}
% \end{figure}

% Percy figures

\begin{figure}
	\includegraphics[width=\columnwidth]{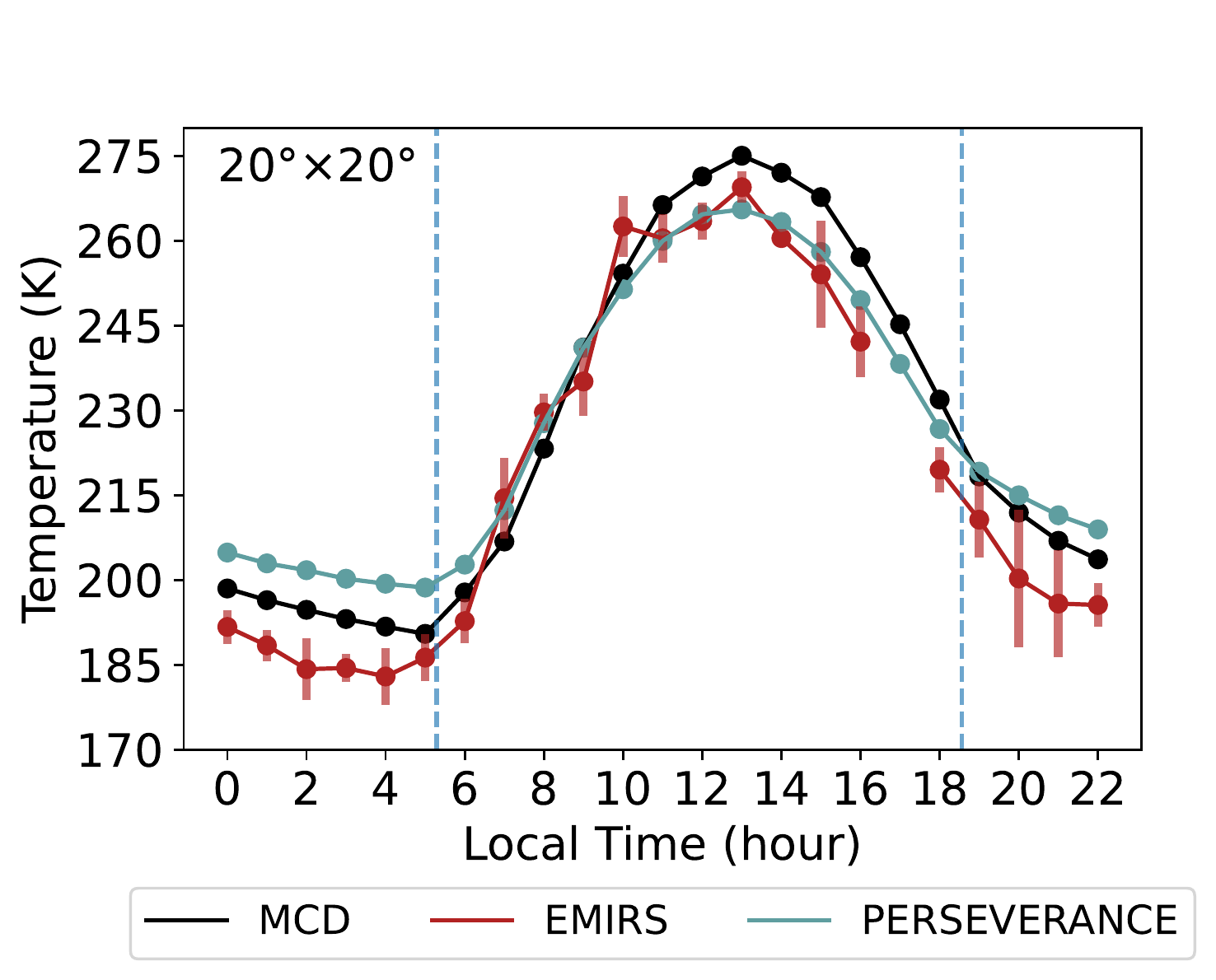}\\
	\includegraphics[width=\columnwidth]{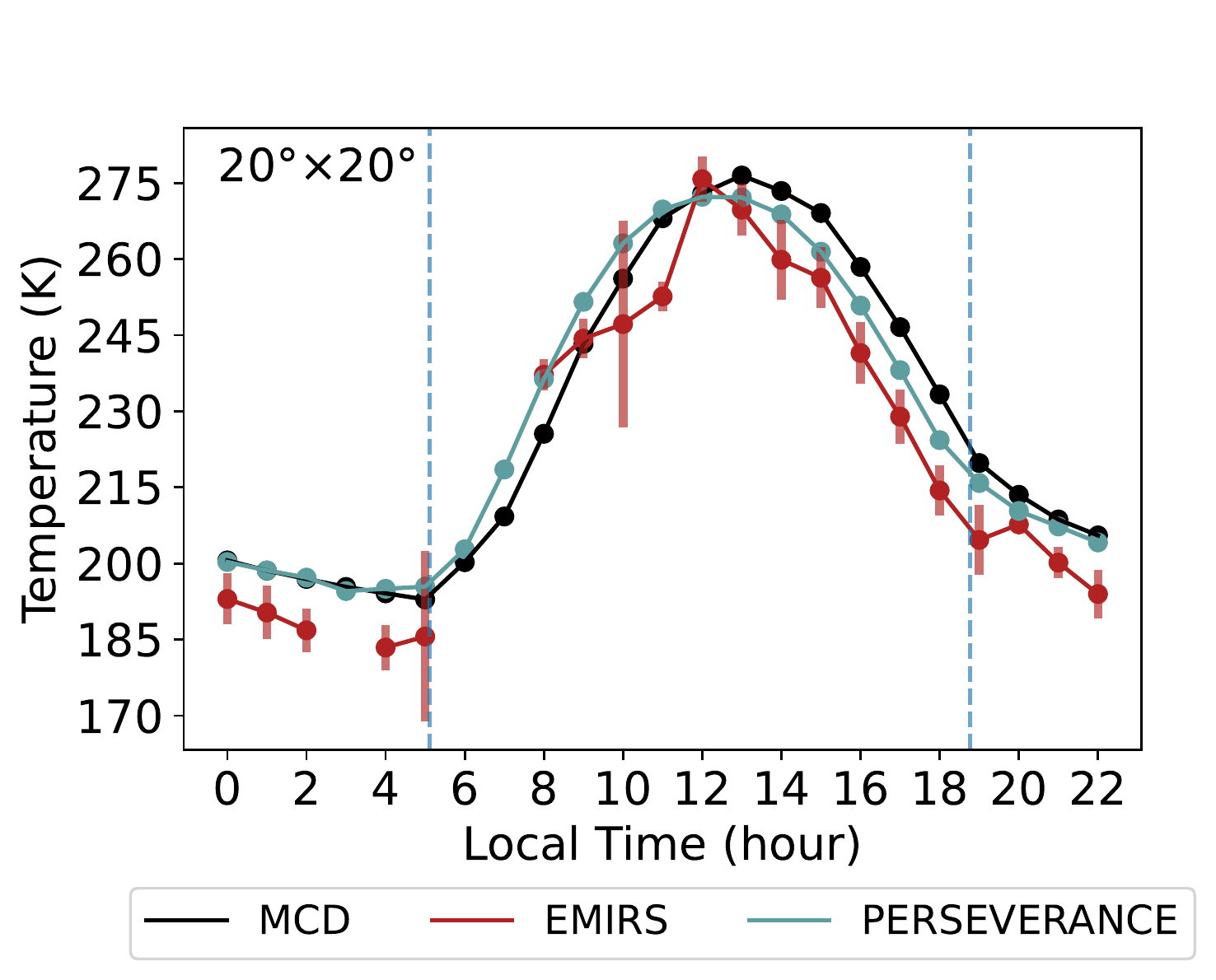}
    \caption{A comparison of average diurnal temperature variation measured by EMIRS/EMM and MEDA/Mars~2020 with the model prediction from MCD. {\it Top panel}: Over $\sim$~30~sols for L$_{s}$ = 65.8-78.9 (July 2021, {\color{red} sols 128-158}). {\it Bottom panel}: Over $\sim$~30~sols for L$_{s}$ = 79.3-92.5 (August 2021, {\color{red} sols 158-188}). The error bars indicate 3-$\sigma$ standard errors. Local dawn and dusk are marked by vertical lines. }
    \label{fig:emirs_perseverance}
\end{figure}
% \begin{figure}
% 	\includegraphics[width=\columnwidth]{SurfJuly031_EMIRS_percy_MCD_3stderr-2.pdf}
%     \caption{A comparison of average diurnal temperature variation over 31 sols measured by EMIRS/EMM and MEDA/Mars 2020 from L$_{s}$ = 65.8 (July 2021) with the model prediction from MCD. The error bars indicate 3-$\sigma$ standard errors. Local dawn and dusk are marked by vertical lines.}
%     \label{fig:emirs_perseverance_july}
% \end{figure}
% Figure \ref{fig:mcd_july} shows the diurnal temperature variation over the month of July 2021.

% \begin{figure}
% 	\includegraphics[width=\columnwidth]{SurfAugust031_EMIRS_percy_MCD_3stderr.pdf}
%     \caption{A comparison of average diurnal temperature variation over 31 sols measured by EMIRS/EMM and MEDA/Mars 2020 L$_{s}$ = 79.3 (August 2021) with the model prediction from MCD. The error bars indicate 3-$\sigma$ standard errors. Local dawn and dusk are marked by vertical lines.}
%     \label{fig:emirs_perseverance_august}
% \end{figure}
Figure~\ref{fig:surface_hours_comparisons} shows the hourly surface temperature measurement comparisons from REMS and EMIRS at the MSL rover site over (Gale crater) $\sim$~30~sols for L$_{s}$ = 65.8-78.9 (July 2021) and 79.3-92.5 (August 2021) respectively. We sample the EMIRS temperature measurements from $2\degree \times 2\degree$, $3\degree \times 3\degree$, $5\degree \times 5\degree$, $7\degree \times 7\degree$, $10\degree \times 10\degree$, $15\degree \times 15\degree$ and $20\degree \times 20\degree$ region centered at the MSL location. Wherever EMIRS observations are available, the peak temperatures show a good match, while the lowest temperatures recorded by REMS are slightly higher.  Figure~ \ref{fig:emirs_curiosity} shows the corresponding average diurnal temperatures, along with the temperatures predicted by the MCD model at the MSL rover site. The average diurnal variation pattern is identical in both EMIRS and REMS temperature measurements and consistent with the MCD model predictions. The average daily temperatures start rising at 05 hrs and reach a peak value of approximately 250K at around 13 hrs in all cases. The EMIRS temperature measurements fall more rapidly after the peak and, in general, the afternoon, evening and night time temperatures are markedly lower compared to measurements by REMS and the MCD model. The temperatures recorded by REMS are generally consistent with the MCD model predictions for the morning and the afternoon. However, they are systematically higher than the MCD model predictions for the night time. The average lowest temperatures vary significantly and are 170~K, 180~K, and 190~K for EMIRS, MCD and REMS respectively.

%The peak temperature all cases is at around 1 pm, and the values are consistent at the hottest time of the day. However, at lower temperatures, EMIRS measurements are significantly lower than both MSL measurements and the model. MSL measurements are higher compared to the model except during the hottest time of the day.

%MSL and MCD are consistent around noon time. EMIRS is systematically lower. MSL is higher compared to both MCD and EMIRS. MSL is higher than MCD during colder parts of the day. 

\subsection{EMM, Mars 2020 and MCD}
%M2020 comments
Figure~\ref{fig:surface_hours_comparisons} shows hourly surface temperature measurement comparisons from MEDA and EMIRS at the Perseverance rover site (Jezero crater) over $\sim$~30~sols for L$_{s}$ = 65.8-78.9 (July 2021) and 79.3-92.5 (August 2021) respectively. Compared to EMIRS, the Perseverance temperature measurements show low maximum to minimum daily temperature variability, and this variability is particularly low for sols 10-25 in July. Figure~\ref{fig:emirs_perseverance} shows the corresponding average diurnal temperature variation, along with the temperatures predicted by the MCD model at the Perseverance rover site. The average diurnal variation pattern is identical in both the EMIRS and the Curiosity temperature measurements and consistent with the MCD model predictions. The average daily temperatures start rising from 04 hrs and reach a peak value at around 13 hrs in all cases. The peak temperature values recorded by EMIRS and Perseverance are approximately equal and are  $\sim$ 10~K lower than the MCD model peak temperature of $\sim$ 270~K. The MCD model predicts afternoon temperatures that are systematically higher by $\sim$ 10K compared to EMIRS, with temperatures obtained by Perseverance falling in between. The night-time temperatures recorded by EMIRS remain lower compared to the MCD model, whereas the Perseverance temperature measurements are slightly higher. The lowest temperatures recorded by EMIRS, MCD, and Perseverance are $\sim$ 180~K, $\sim$ 190~K, and $\sim$ 200~K respectively. Appendix figures C2-C5 show hourly EMIRS surface temperature comparisons with Curiosity and Perseverance observations for other months as well as smaller field of views (FOV) of EMIRS about the rover locations. Appendix figures D2-D5 show corresponding comparisons for average diurnal temperatures.

% 20 by 20 degrees. 
% EMIRS is lower at night compared to both MCD and M2020. M2020 is higher until around 10 am compared to MCD but is consistent at night. There is no M2020 data in February but EMIRS is low compared to MCD except at 7-8 hours.  

\section{Conclusion and Discussion}
EMM observations, for the first time have enabled us to study the surface temperatures at all local times for most locations on Mars. This has enabled us to compare EMM/EMIRS observations taken from the orbit at overlapping times with surface rovers MSL and Mars 2020. We find that the overall trend in daily temperature variability is consistent between different missions and also with MCD estimates. Hourly temperature variability of EMIRS is in good agreement with MSL. However, EMIRS measurements are systematically lower, especially at night-times at lowest temperatures. There could be several reasons leading to this discrepancy in measurements. One potential issue with this analysis is the vast difference between the FOV of EMIRS (as discussed earlier) and rovers. Although measurements from rovers are sensitive to variability arising from both small-scale local and large-scale regional features, EMIRS is sensitive to only large-scale regional features. Within the FOV of EMIRS, a number of geological units with varying thermal inertia and surface roughness could exist (see appendix figures E1-E3). Different thermal inertia of rocks, dust and sand would result in uneven rate of change of temperature, influencing the diurnal temperature variation. Also, brightness temperature obtained from observations is highly sensitive to rock abundance, especially at night \citep{emirswolfe}. Preliminary work by \cite{emirswolfe} found that EMIRS observations are sensitive to rock abundance, especially near the equator and mid-latitude regions. The surface scattering and emission also depends on the roughness of the surface, which in turn would lead to differences in estimated values of surface temperature. The seemingly odd kinks in EMIRS averaged data (Figures~ \ref{fig:emirs_curiosity} and ~\ref{fig:emirs_perseverance}) are an artefact of the non-uniform spatial sampling from the larger FOV, and go away at smaller FOV as seen in figures D4 and D5 in the appendix. The corresponding number of observations are lower and error bars are larger, and hence these apparent peaks are not significant for the diurnal temperature variation trends discussed.

EMIRS observations with smaller grid size are taken in a similar thermal environment as rovers averaged over a period of time. Being an instrument on board an orbiter, having a low spatial resolution is a limitation of EMIRS. With more observations made by the orbiter, we hope to create a better picture of the thermal environment of Mars, which will also enable better comparison with rover measurements. More work needs to be done in order to disentangle the influence of regional versus local features, between different rock units and other factors such as surface roughness which might influence the estimates of surface temperature, and this analysis is an important first step in that direction. Additionally, estimates from climate models too vary significantly, for example, up to 10 K difference in surface temperature was seen across nine different models in the Jezero crater region \citep{newman2021multi}. With this work of comparing observations made from the orbit with those from the surface, we hope to better constrain models leading to advancements in weather and climate modeling, especially model-to-model differences. We plan to compare seasonal changes observed between different missions though one Mars-year, which is also the duration of EMM's primary mission. This effort will also assist with improved planning of future robotic missions \citep{fonseca2019marswrf} as well as crewed missions to the planet \citep{drake2010human}.

\section*{Data Availability}
EMM data was obtained from the EMM Science Data Center (SDC), and MSL and Mars 2020 data from NASA's Planetary Data System (PDS), both are open access repositories.
\section*{Acknowledgments}

 This work was supported by the New York University Abu Dhabi (NYUAD) Institute Research Grant G1502 and the ASPIRE Award for Research Excellence (AARE) Grant S1560 by the Advanced Technology Research Council (ATRC). This work utilized the High Performance Computing (HPC) resources of NYUAD. We thank Prof. K. R. Sreenivasan for his constant encouragement and support for the project.

%%%%%%%%%%%%%%%%%%%%%%%%%%%%%%%%%%%%%%%%%%%%%%%%%%



\bibliographystyle{mnras}
\bibliography{example} % if your bibtex file is called example.bib

\begin{thebibliography}{}
\makeatletter
\relax
\def\mn@urlcharsother{\let\do\@makeother \do\$\do\&\do\#\do\^\do\_\do\%\do\~}
\def\mn@doi{\begingroup\mn@urlcharsother \@ifnextchar [ {\mn@doi@}
  {\mn@doi@[]}}
\def\mn@doi@[#1]#2{\def\@tempa{#1}\ifx\@tempa\@empty \href
  {http://dx.doi.org/#2} {doi:#2}\else \href {http://dx.doi.org/#2} {#1}\fi
  \endgroup}
\def\mn@eprint#1#2{\mn@eprint@#1:#2::\@nil}
\def\mn@eprint@arXiv#1{\href {http://arxiv.org/abs/#1} {{\tt arXiv:#1}}}
\def\mn@eprint@dblp#1{\href {http://dblp.uni-trier.de/rec/bibtex/#1.xml}
  {dblp:#1}}
\def\mn@eprint@#1:#2:#3:#4\@nil{\def\@tempa {#1}\def\@tempb {#2}\def\@tempc
  {#3}\ifx \@tempc \@empty \let \@tempc \@tempb \let \@tempb \@tempa \fi \ifx
  \@tempb \@empty \def\@tempb {arXiv}\fi \@ifundefined
  {mn@eprint@\@tempb}{\@tempb:\@tempc}{\expandafter \expandafter \csname
  mn@eprint@\@tempb\endcsname \expandafter{\@tempc}}}

\bibitem[\protect\citeauthoryear{Almatroushi et~al.,}{Almatroushi
  et~al.}{2021}]{almatroushi2021emirates}
Almatroushi H.,  et~al., 2021, Space Science Reviews, 217, 1

\bibitem[\protect\citeauthoryear{Amiri et~al.,}{Amiri
  et~al.}{2022}]{amiri2022emirates}
Amiri H.,  et~al., 2022, Space Science Reviews, 218, 1

\bibitem[\protect\citeauthoryear{Bhardwaj}{Bhardwaj}{2014}]{bhardwaj2014indian}
Bhardwaj A.,  2014, 40th COSPAR Scientific Assembly, 40, C3

\bibitem[\protect\citeauthoryear{Christensen et~al.,}{Christensen
  et~al.}{2001}]{christensen2001mars}
Christensen P.~R.,  et~al., 2001, Journal of Geophysical Research: Planets,
  106, 23823

\bibitem[\protect\citeauthoryear{Drake, Hoffman  \& Beaty}{Drake
  et~al.}{2010}]{drake2010human}
Drake B.~G.,  Hoffman S.~J.,   Beaty D.~W.,  2010, in 2010 IEEE Aerospace
  Conference. pp 1--24

\bibitem[\protect\citeauthoryear{Edwards et~al.,}{Edwards
  et~al.}{2018}]{edwards2018thermophysical}
Edwards C.~S.,  et~al., 2018, Journal of Geophysical Research: Planets, 123,
  1307

\bibitem[\protect\citeauthoryear{Edwards et~al.,}{Edwards
  et~al.}{2021}]{edwards2021emirates}
Edwards C.~S.,  et~al., 2021, Space science reviews, 217, 1

\bibitem[\protect\citeauthoryear{Fonseca, Zorzano  \&
  Mart{\'\i}n-Torres}{Fonseca et~al.}{2019}]{fonseca2019marswrf}
Fonseca R.~M.,  Zorzano M.-P.,   Mart{\'\i}n-Torres J.,  2019, Earth and Space
  Science, 6, 1440

\bibitem[\protect\citeauthoryear{Formisano et~al.,}{Formisano
  et~al.}{2005}]{formisano2005planetary}
Formisano V.,  et~al., 2005, Planetary and Space Science, 53, 963

\bibitem[\protect\citeauthoryear{G{\'o}mez-Elvira et~al.,}{G{\'o}mez-Elvira
  et~al.}{2012}]{gomez2012rems}
G{\'o}mez-Elvira J.,  et~al., 2012, Space science reviews, 170, 583

\bibitem[\protect\citeauthoryear{He, Arvidson, O’Sullivan, Morris, Condus,
  Hughes  \& Powell}{He et~al.}{2022}]{he2022surface}
He L.,  Arvidson R.,  O’Sullivan J.,  Morris R.,  Condus T.,  Hughes M.,
  Powell K.,  2022, Journal of Geophysical Research: Planets, p. e2021JE007092

\bibitem[\protect\citeauthoryear{Hess, Henry, Leovy, Ryan  \& Tillman}{Hess
  et~al.}{1977}]{hess1977meteorological}
Hess S.,  Henry R.,  Leovy C.~B.,  Ryan J.,   Tillman J.~E.,  1977, Journal of
  Geophysical Research, 82, 4559

\bibitem[\protect\citeauthoryear{Jakosky}{Jakosky}{2021}]{jakosky2021atmospheric}
Jakosky B.~M.,  2021, Annual Review of Earth and Planetary Sciences, 49, 71

\bibitem[\protect\citeauthoryear{Korablev et~al.,}{Korablev
  et~al.}{2018}]{korablev2018atmospheric}
Korablev O.,  et~al., 2018, Space Science Reviews, 214, 1

\bibitem[\protect\citeauthoryear{Mart{\'\i}nez et~al.,}{Mart{\'\i}nez
  et~al.}{2017}]{martinez2017modern}
Mart{\'\i}nez G.,  et~al., 2017, Space Science Reviews, 212, 295

\bibitem[\protect\citeauthoryear{Mart{\'\i}nez et~al.,}{Mart{\'\i}nez
  et~al.}{2021}]{martinez2021surface}
Mart{\'\i}nez G.,  et~al., 2021, Journal of Geophysical Research: Planets, 126,
  e2020JE006804

\bibitem[\protect\citeauthoryear{Martinez et~al.,}{Martinez
  et~al.}{2022}]{martinez2022thermal}
Martinez G.,  et~al., 2022, LPI Contributions, 2678, 2024

\bibitem[\protect\citeauthoryear{Martínez}{Martínez}{2022}]{martinezp}
Martínez G.,  2022, Personal communication

\bibitem[\protect\citeauthoryear{Mason \& Smith}{Mason \&
  Smith}{2021}]{mason2021temperature}
Mason E.~L.,  Smith M.~D.,  2021, Icarus, 360, 114350

\bibitem[\protect\citeauthoryear{Millour, Forget, Spiga, Vals, Zakharov  \&
  Montabone}{Millour et~al.}{2018}]{millour2018mars}
Millour E.,  Forget F.,  Spiga A.,  Vals M.,  Zakharov V.,   Montabone L.,
  2018, in , From Mars Express to ExoMars, 27-28 February 2018, Madrid, Spain

\bibitem[\protect\citeauthoryear{Newman et~al.,}{Newman
  et~al.}{2021}]{newman2021multi}
Newman C.,  et~al., 2021, Space science reviews, 217, 1

\bibitem[\protect\citeauthoryear{Newman et~al.,}{Newman
  et~al.}{2022}]{newman2022dynamic}
Newman C.~E.,  et~al., 2022, Science Advances, 8, eabn3783

\bibitem[\protect\citeauthoryear{Petrosyan et~al.,}{Petrosyan
  et~al.}{2011}]{petrosyan2011martian}
Petrosyan A.,  et~al., 2011, Reviews of Geophysics, 49

\bibitem[\protect\citeauthoryear{Piqueux et~al.,}{Piqueux
  et~al.}{2021}]{piqueux2021soil}
Piqueux S.,  et~al., 2021, Journal of Geophysical Research: Planets, 126,
  e2021JE006859

\bibitem[\protect\citeauthoryear{Pla-Garc{\'\i}a et~al.,}{Pla-Garc{\'\i}a
  et~al.}{2020}]{pla2020meteorological}
Pla-Garc{\'\i}a J.,  et~al., 2020, Space science reviews, 216, 1

\bibitem[\protect\citeauthoryear{Rodriguez-Manfredi et~al.,}{Rodriguez-Manfredi
  et~al.}{2021}]{rodriguez2021mars}
Rodriguez-Manfredi J.~A.,  et~al., 2021, Space science reviews, 217, 1

\bibitem[\protect\citeauthoryear{Sagan \& Mullen}{Sagan \&
  Mullen}{1972}]{sagan1972earth}
Sagan C.,  Mullen G.,  1972, Science, 177, 52

\bibitem[\protect\citeauthoryear{Sebasti{\'a}n, Armiens, G{\'o}mez-Elvira,
  Zorzano, Martinez-Frias, Esteban  \& Ramos}{Sebasti{\'a}n
  et~al.}{2010}]{sebastian2010rover}
Sebasti{\'a}n E.,  Armiens C.,  G{\'o}mez-Elvira J.,  Zorzano M.~P.,
  Martinez-Frias J.,  Esteban B.,   Ramos M.,  2010, Sensors, 10, 9211

\bibitem[\protect\citeauthoryear{Sebasti{\'a}n, Mart{\'\i}nez, Ramos,
  P{\'e}rez-Grande, Sobrado  \& Manfredi}{Sebasti{\'a}n
  et~al.}{2021}]{sebastian2021thermal}
Sebasti{\'a}n E.,  Mart{\'\i}nez G.,  Ramos M.,  P{\'e}rez-Grande I.,  Sobrado
  J.,   Manfredi J. A.~R.,  2021, Acta Astronautica, 182, 144

\bibitem[\protect\citeauthoryear{Spanovich, Smith, Smith, Wolff, Christensen
  \& Squyres}{Spanovich et~al.}{2006}]{spanovich2006surface}
Spanovich N.,  Smith M.,  Smith P.,  Wolff M.,  Christensen P.,   Squyres S.,
  2006, Icarus, 180, 314

\bibitem[\protect\citeauthoryear{Spohn et~al.,}{Spohn
  et~al.}{2018}]{spohn2018heat}
Spohn T.,  et~al., 2018, Space Science Reviews, 214, 1

\bibitem[\protect\citeauthoryear{Vago et~al.,}{Vago et~al.}{2015}]{vago2015esa}
Vago J.,  et~al., 2015, Solar System Research, 49, 518

\bibitem[\protect\citeauthoryear{{Wolfe}, {Edwards}, {Smith}, {Christensen},
  {Smith}, {Badri}  \& {Anwar}}{{Wolfe} et~al.}{2022}]{emirswolfe}
{Wolfe} C.~A.,  {Edwards} C.~S.,  {Smith} M.~D.,  {Christensen} P.~R.,  {Smith}
  N.~M.,  {Badri} K.,   {Anwar} S.,  2022, in LPI Contributions. p.~2804

\bibitem[\protect\citeauthoryear{Wolfgang \& Lopez}{Wolfgang \&
  Lopez}{2015}]{wolfgang2015rocky}
Wolfgang A.,  Lopez E.,  2015, The Astrophysical Journal, 806

\bibitem[\protect\citeauthoryear{Zurbuchen}{Zurbuchen}{2017}]{zurbuchen2017mars}
Zurbuchen T.~H.,  2017, Presentation to the National Academies, 28

\makeatother
\end{thebibliography}

%\begin{thebibliography}{99}
%\end{thebibliography}

%%%%%%%%%%%%%%%%%%%%%%%%%%%%%%%%%%%%%%%%%%%%%%%%%%

%%%%%%%%%%%%%%%%% APPENDICES %%%%%%%%%%%%%%%%%%%%%


 %%%%%%%%%%%%%%%%%%%%%%%%%%%%%%%%%%%%%%%%%%%%%%%%%%
%%%%%%%%%%%%%%%%% APPENDICES %%%%%%%%%%%%%%%%%%%%%
 \appendix
 \onecolumn
 \section{EMIRS L2 and L3 Surface Temperature Comparison}
 \makeatletter
\def\@captype{figure}
\makeatother
 \centering
 	\includegraphics[width=0.65\columnwidth]{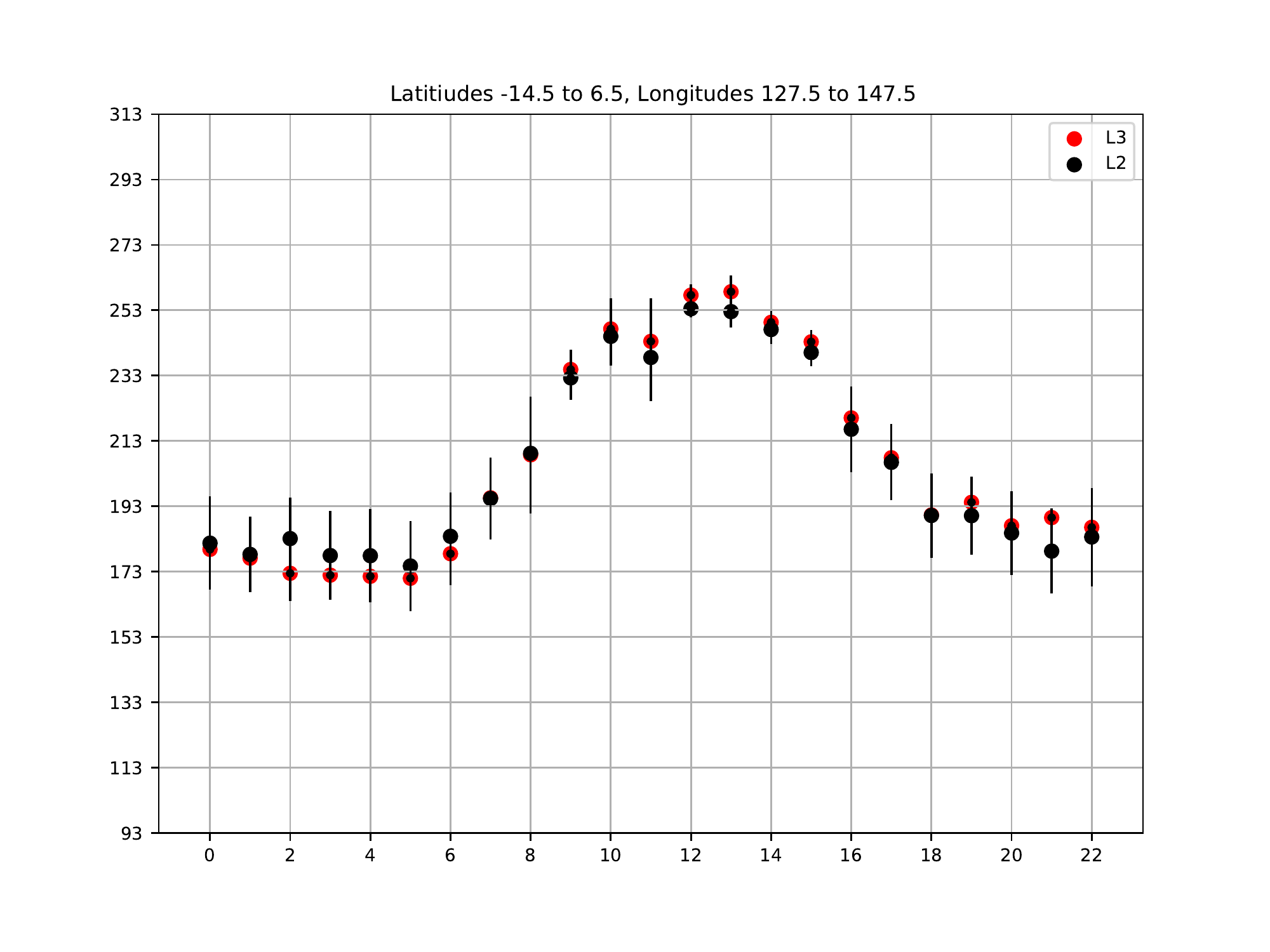}
     \caption{L2 vs L3 surface temperature measurements by EMIRS/EMM during the month of July 2021 (starting L$_{s}$ = 65.8)}
     \label{fig:surface_l2l3_july}

 \newpage
 \section{Global Surface Temperature Measurements}
 \makeatletter
\def\@captype{figure}
\makeatother
 \centering
 	\includegraphics[width=\columnwidth]{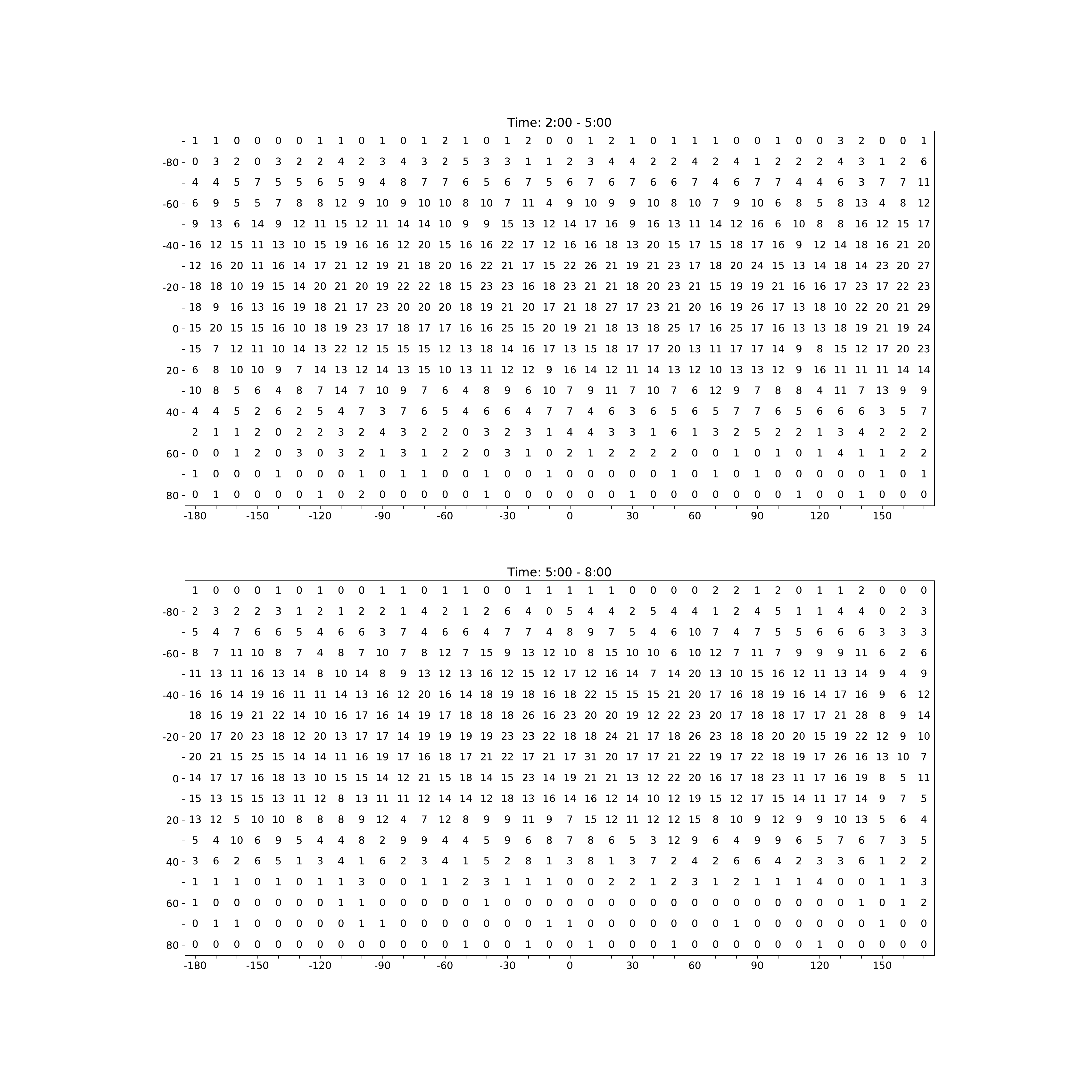}
     \caption{A representative figure showing the number of EMIRS observations available in each spatial bin to calculate the global average surface temperature maps in Figure 1.}
     \label{fig:measurements_bin_july}
 \makeatletter
\def\@captype{figure}
\makeatother
 \centering
 	\includegraphics[width=0.4\columnwidth,trim={0.0cm 73.0cm 0.0cm 0.0cm},clip]{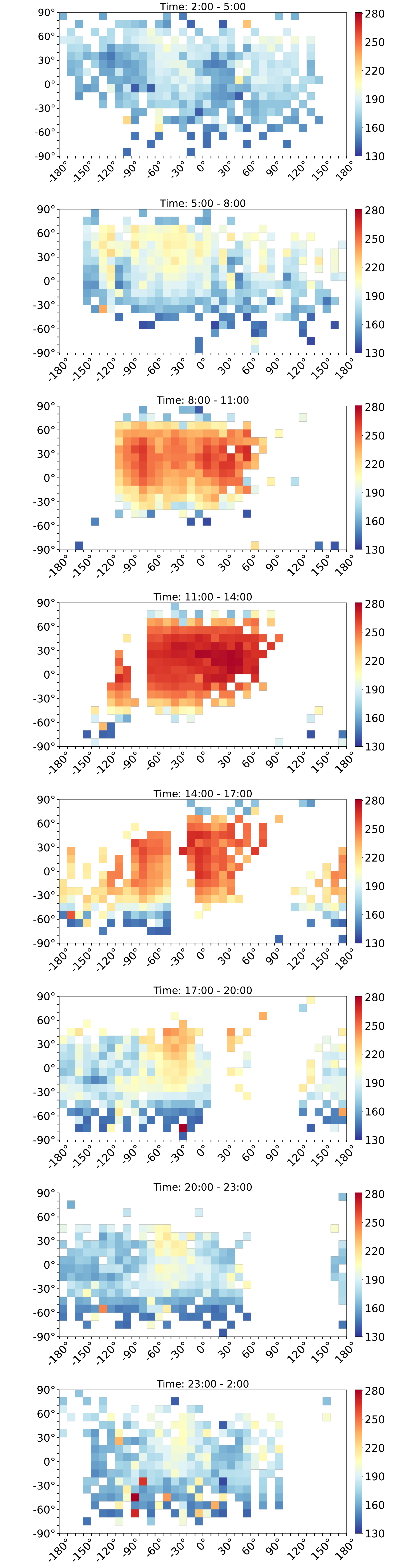}
 	\includegraphics[width=0.4\columnwidth,trim={0.0cm 0.0cm 0.0cm 73.0cm},clip]{Surface_Map_May.pdf}
     \caption{Global surface temperature measurements by EMIRS/EMM during the month of May 2021 (starting L$_{s}$ = 38.8)}
     \label{fig:surface_map_may}
 
 \begin{figure}
 \centering
 	\includegraphics[width=0.4\columnwidth,trim={0.0cm 73.0cm 0.0cm 0.0cm},clip]{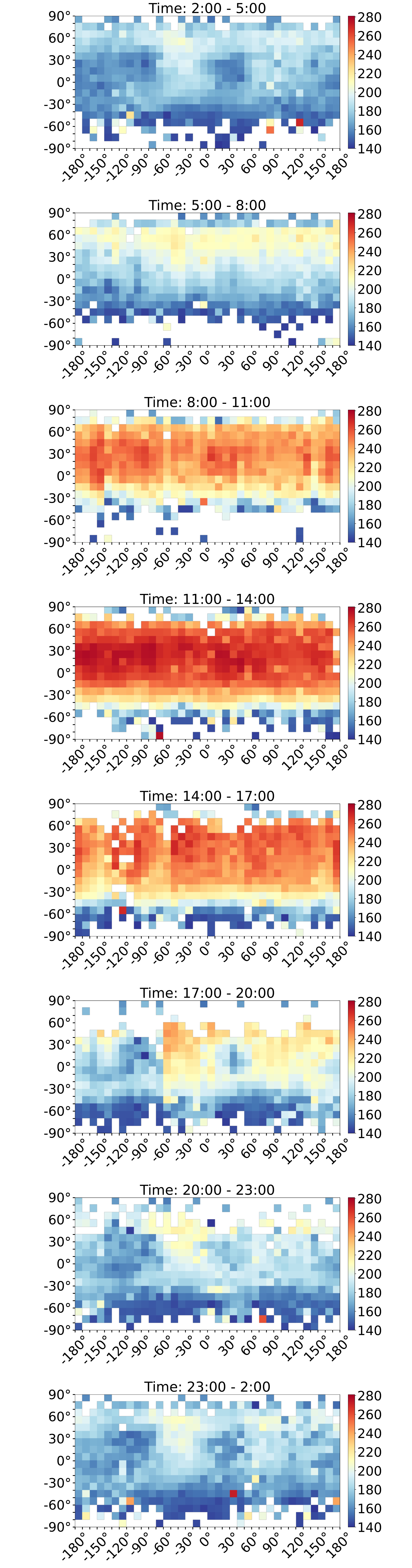}
 	\includegraphics[width=0.4\columnwidth,trim={0.0cm 0.0cm 0.0cm 73.0cm},clip]{Surface_Map_June.pdf}
     \caption{Global surface temperature measurements by EMIRS/EMM during the month of June 2021 (starting L$_{s}$ = 52.6)}
     \label{fig:surface_map_june}
 \end{figure}

 \begin{figure}
 \centering
 	\includegraphics[width=0.4\columnwidth,trim={0.0cm 73.0cm 0.0cm 0.0cm},clip]{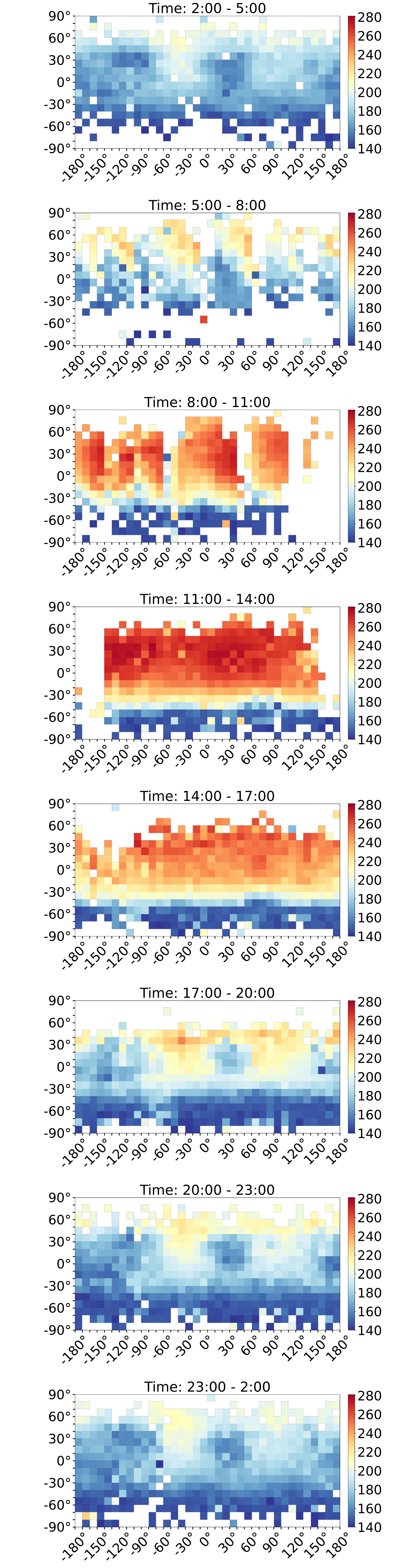}
 	\includegraphics[width=0.4\columnwidth,trim={0.0cm 0.0cm 0.0cm 73.0cm},clip]{Surface_Map_August.pdf}
     \caption{Global surface temperature measurements by EMIRS/EMM during the month of August 2021 (starting L$_{s}$ = 79.3).}
     \label{fig:surface_map_aug}
 \end{figure}
 \newpage
 
 \section{Hourly temperature comparisons}
 \makeatletter
\def\@captype{figure}
\makeatother
 \centering
 	\includegraphics[width=0.6\columnwidth]{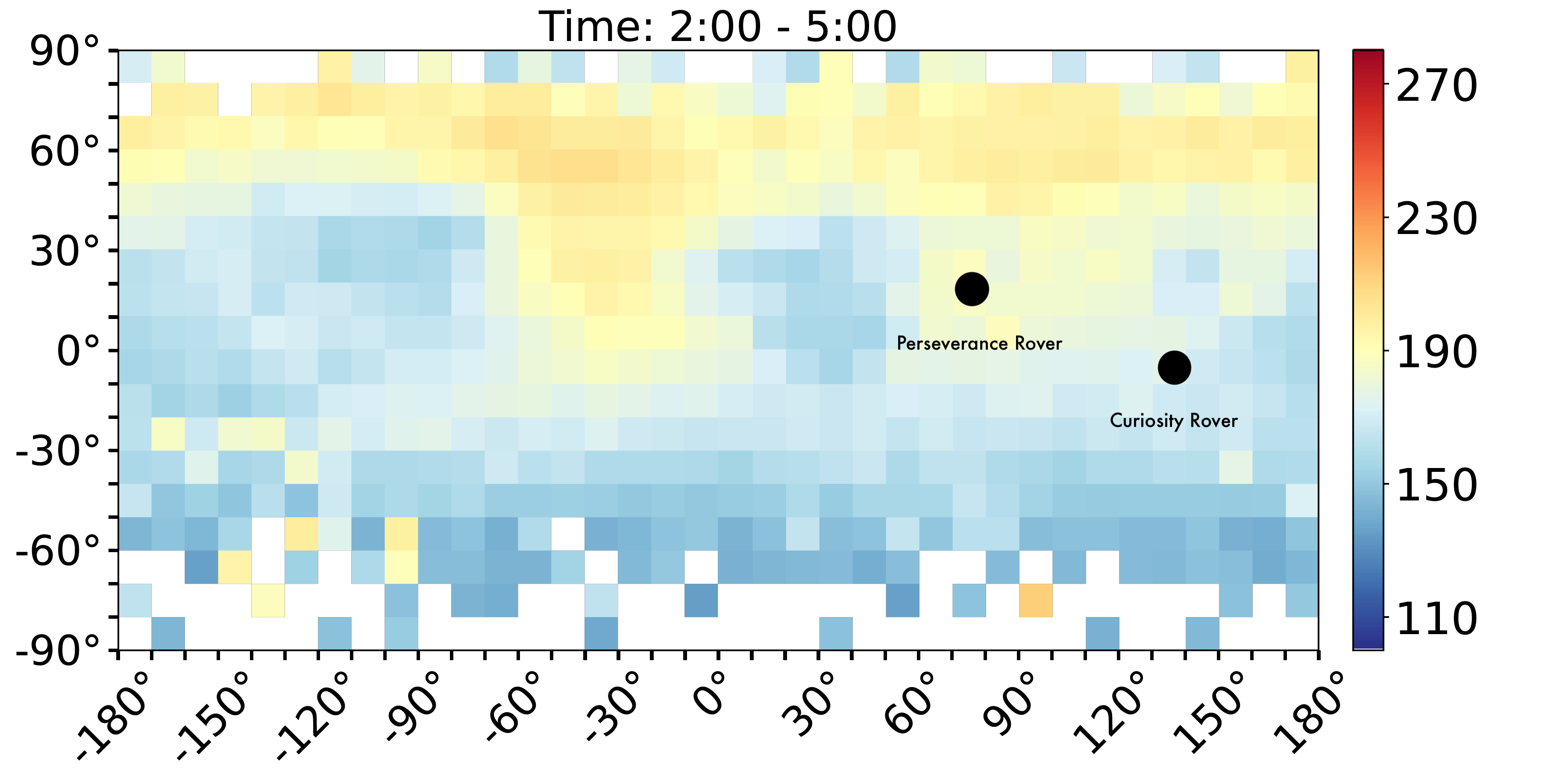}\\
     \caption{The Perseverance and Curiosity roves sites on Mars surface.}
     \label{fig:surface_rovers}
     \vspace{\baselineskip}
 
 \makeatletter
\def\@captype{figure}
\makeatother
 \centering
 	\includegraphics[width=0.65\columnwidth]{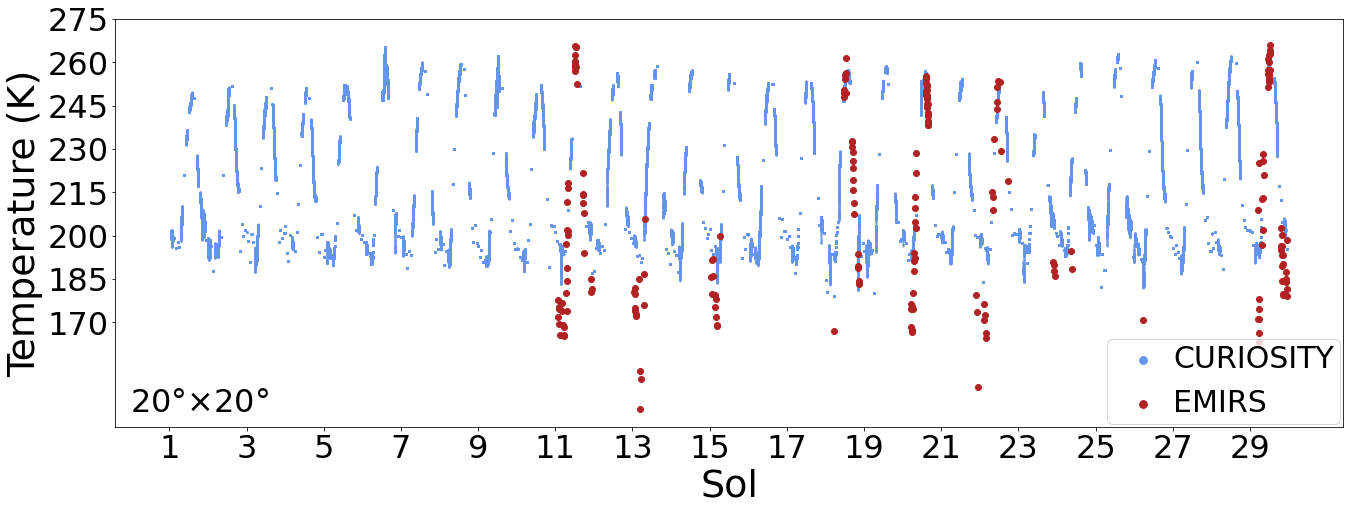}
     \caption{Hourly surface temperature measurement from REMS/MSL and EMIRS starting L$_{s}$ = 52.6 (June 2021)}
     \label{fig:surface_hours_msl_june}
     \vspace{\baselineskip}
 \makeatletter
\def\@captype{figure}
\makeatother
 \centering
 	\includegraphics[width=0.65\columnwidth]{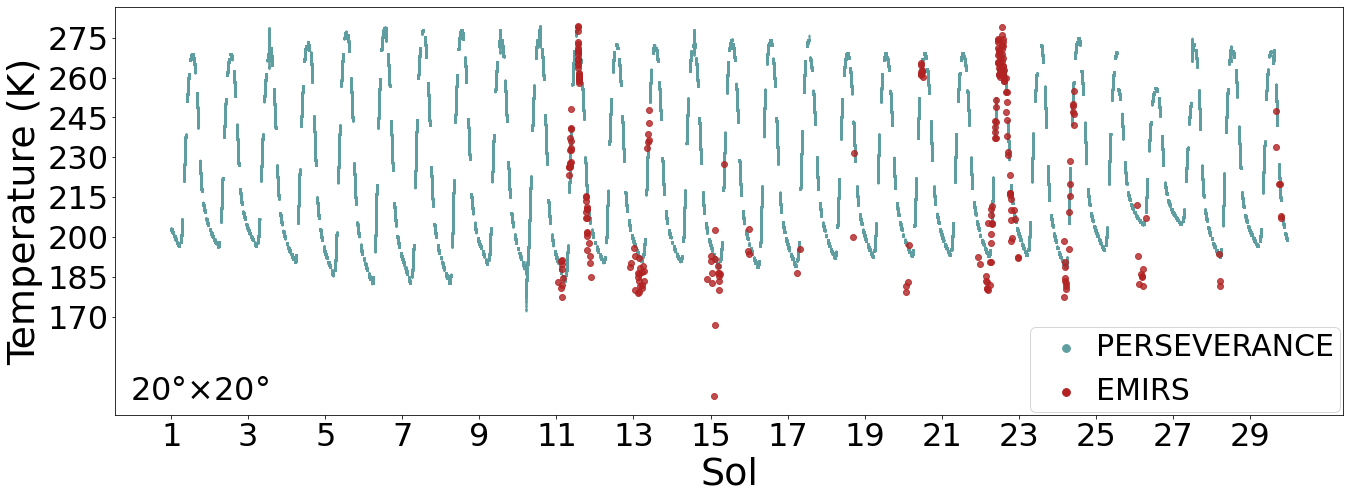}
     \caption{Hourly surface temperature measurement from MEDA/M2020 and EMIRS starting L$_{s}$ = 52.6 (June 2021)}
     \label{fig:surface_hours_percy_june}
  
  \begin{figure}
 	\includegraphics[width=0.5\columnwidth]{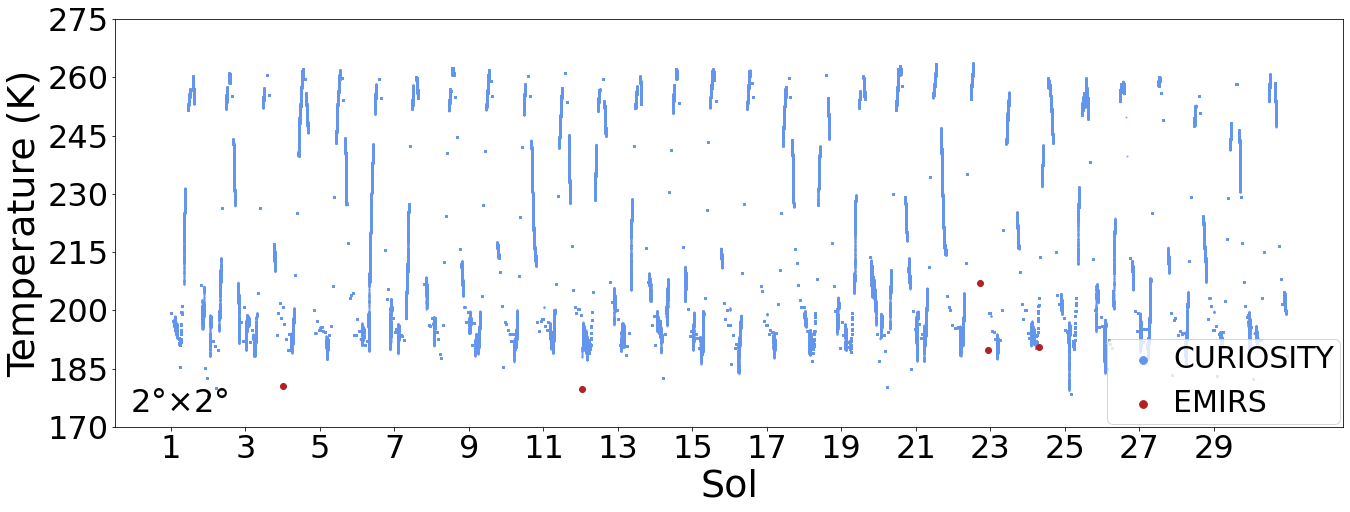}
 	 	\includegraphics[width=0.5\columnwidth]{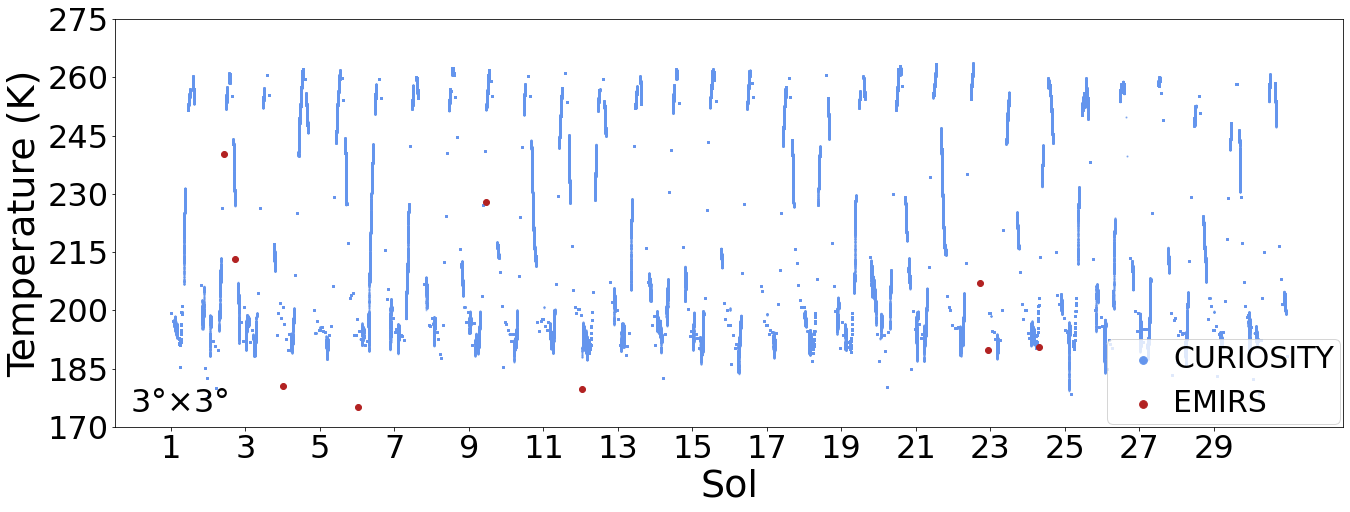}
 	\includegraphics[width=0.5\columnwidth]{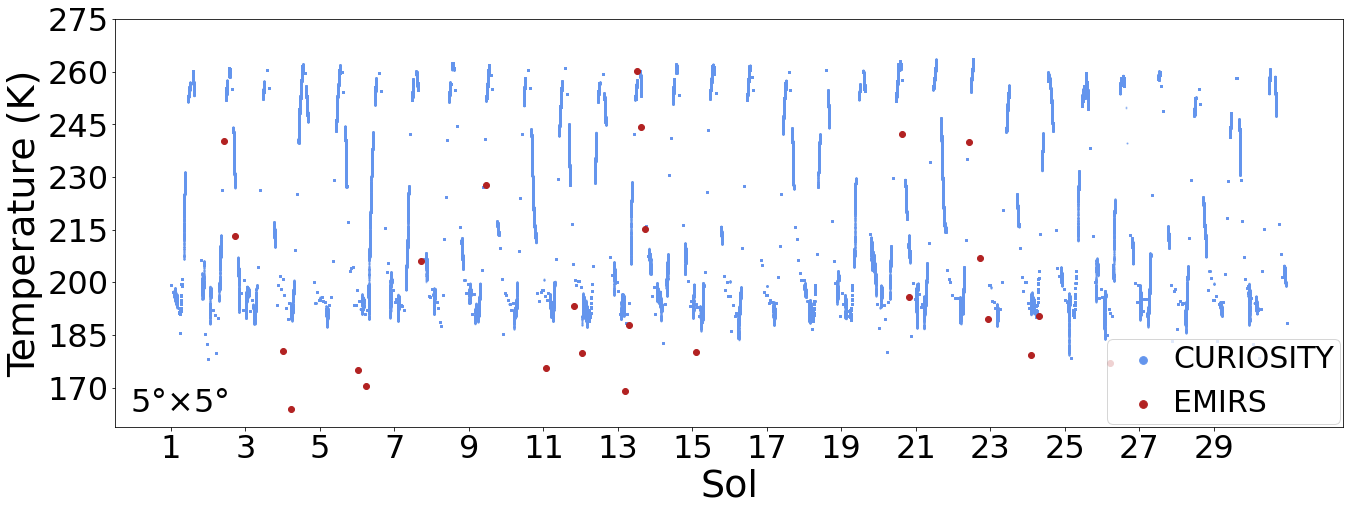}
 	\includegraphics[width=0.5\columnwidth]{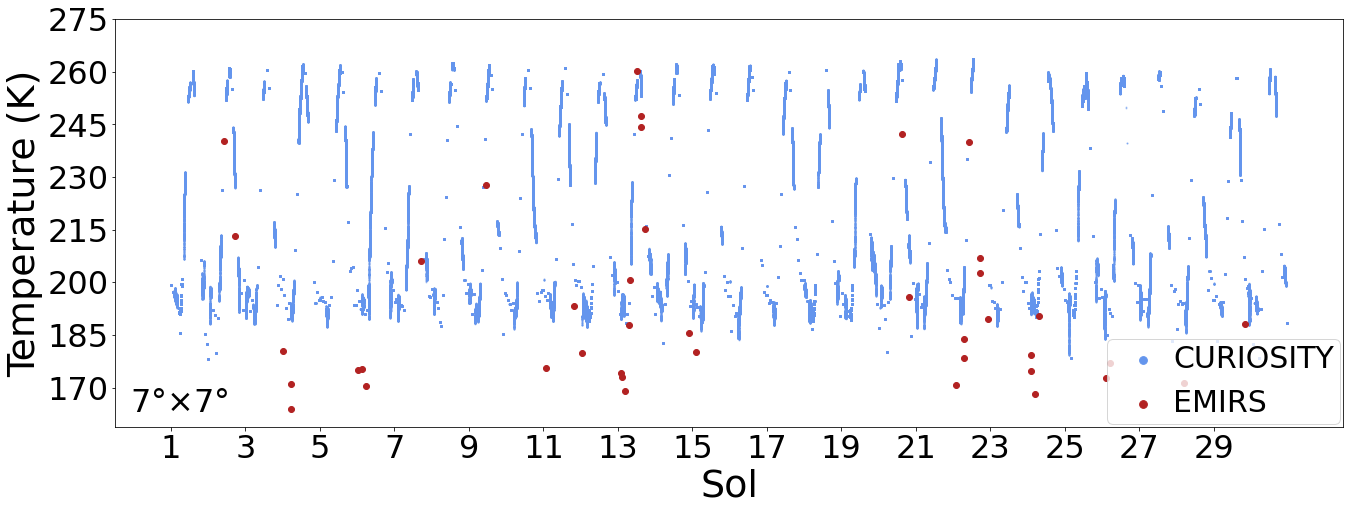}
 	 	\includegraphics[width=0.5\columnwidth]{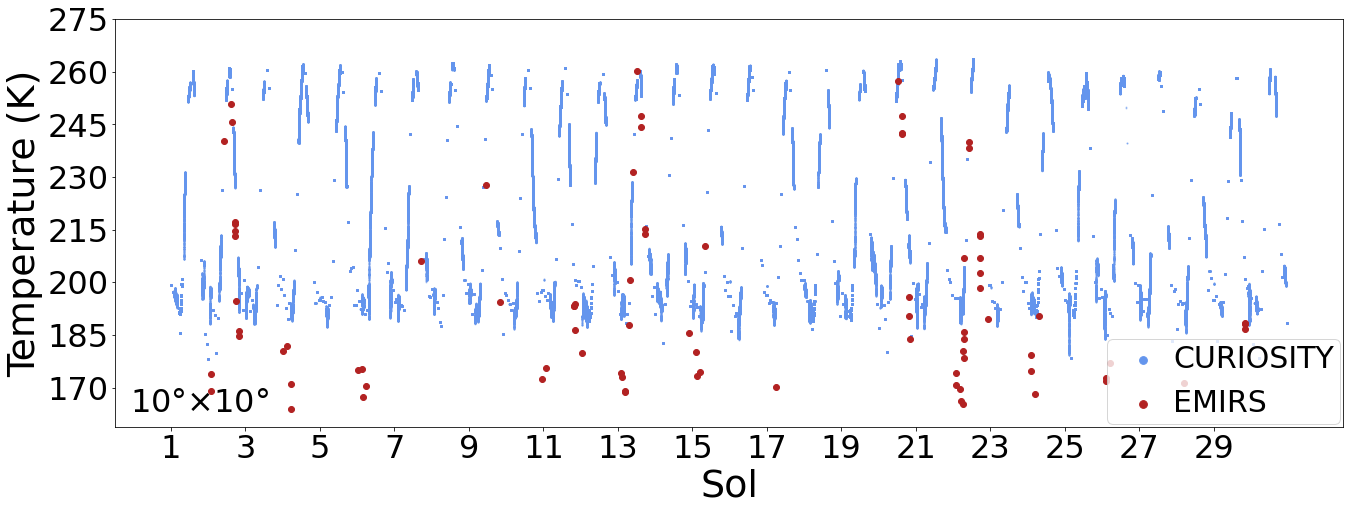}
 	\includegraphics[width=0.5\columnwidth]{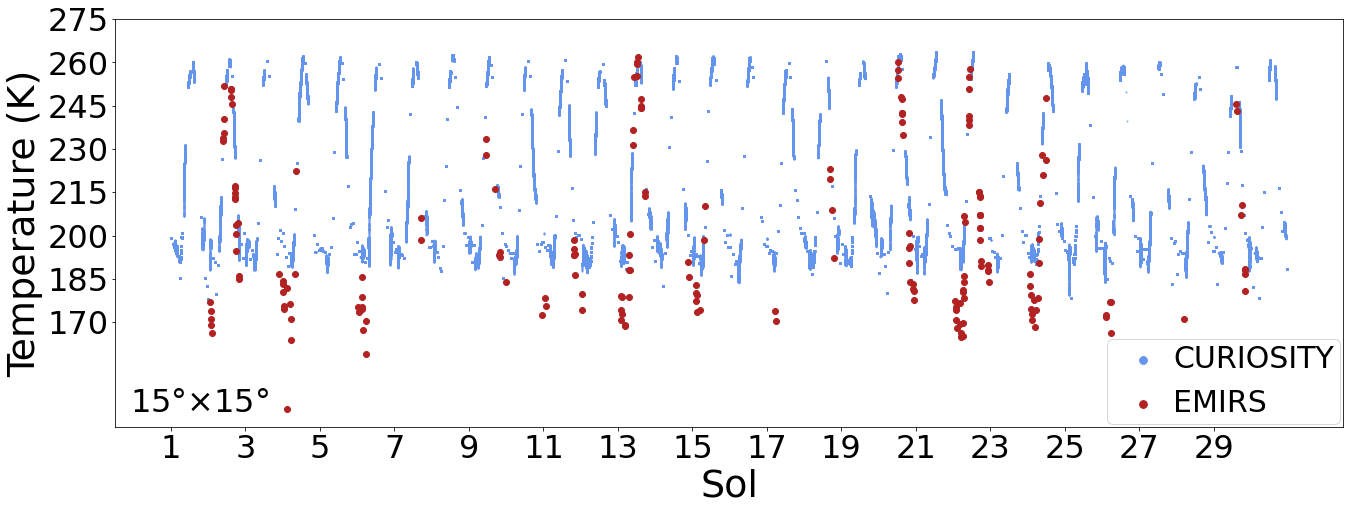}

     \caption{Comparison of hourly Curiosity and EMIRS surface temperature observations of various grid sizes --- $2\degree \times 2\degree$ ({\it top left}), $3\degree \times 3\degree$ ({\it top right}), $5\degree \times 5\degree$ ({\it middle left}), $7\degree \times 7\degree$ ({\it middle right}), $10\degree \times 10\degree$ ({\it bottom left}), $15\degree \times 15\degree$ ({\it  bottom right}) in July 2021.}
     \label{fig:comparison_grid3}
 \end{figure}
 
  \begin{figure}
 	\includegraphics[width=0.5\columnwidth]{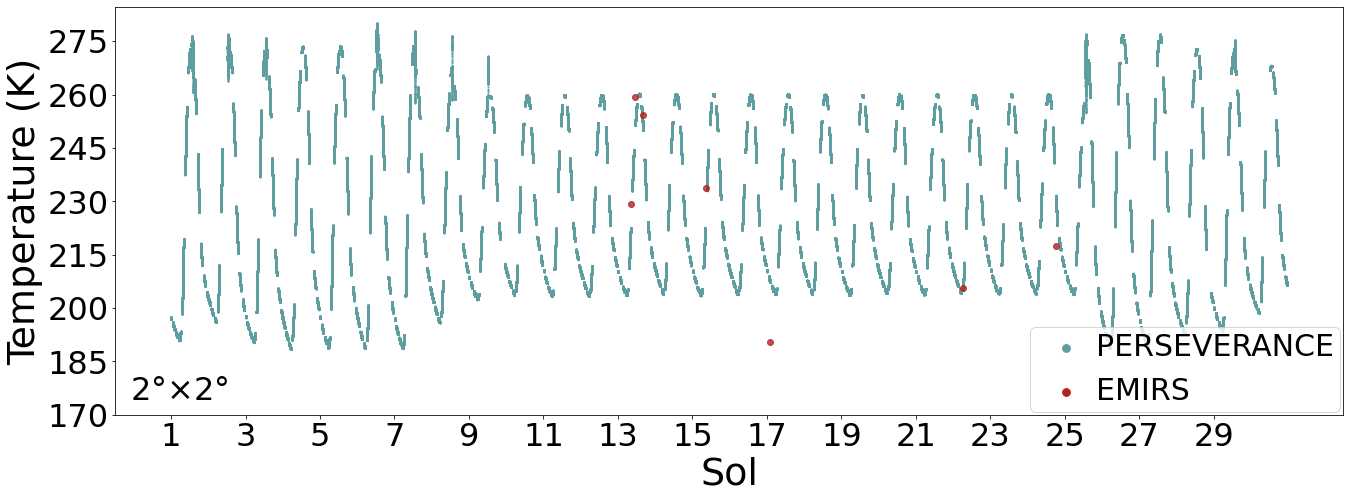}
 	 	\includegraphics[width=0.5\columnwidth]{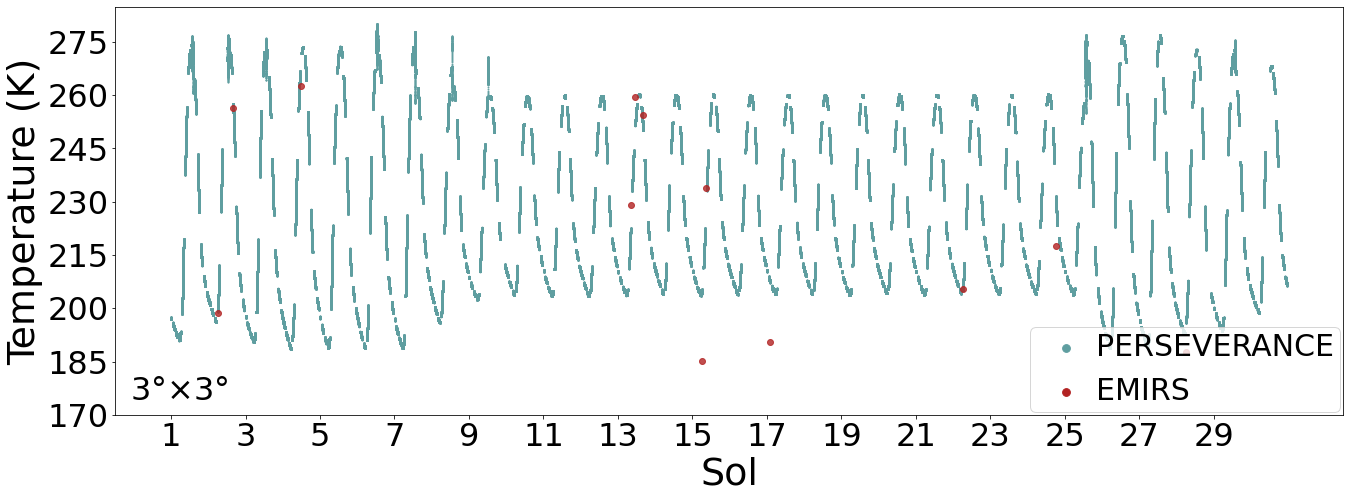}
 	\includegraphics[width=0.5\columnwidth]{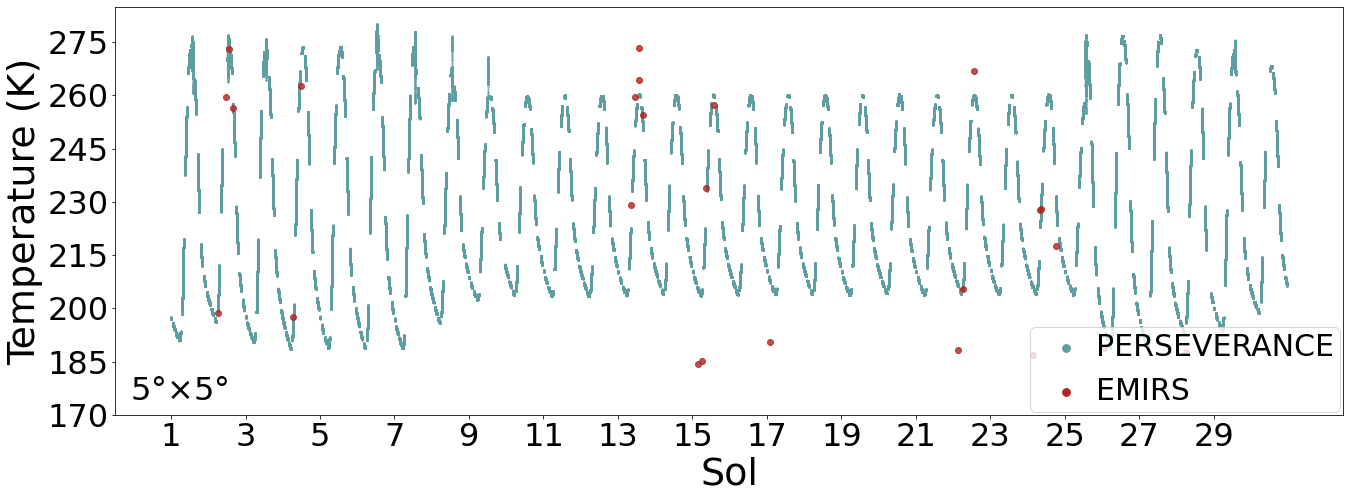}
 	\includegraphics[width=0.5\columnwidth]{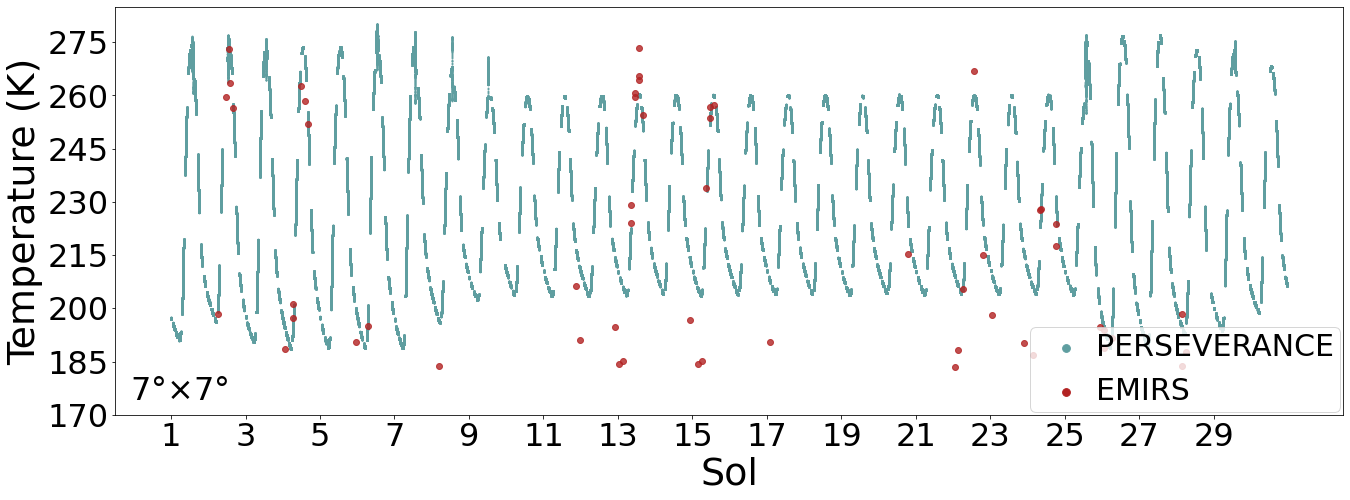}
 	 	\includegraphics[width=0.5\columnwidth]{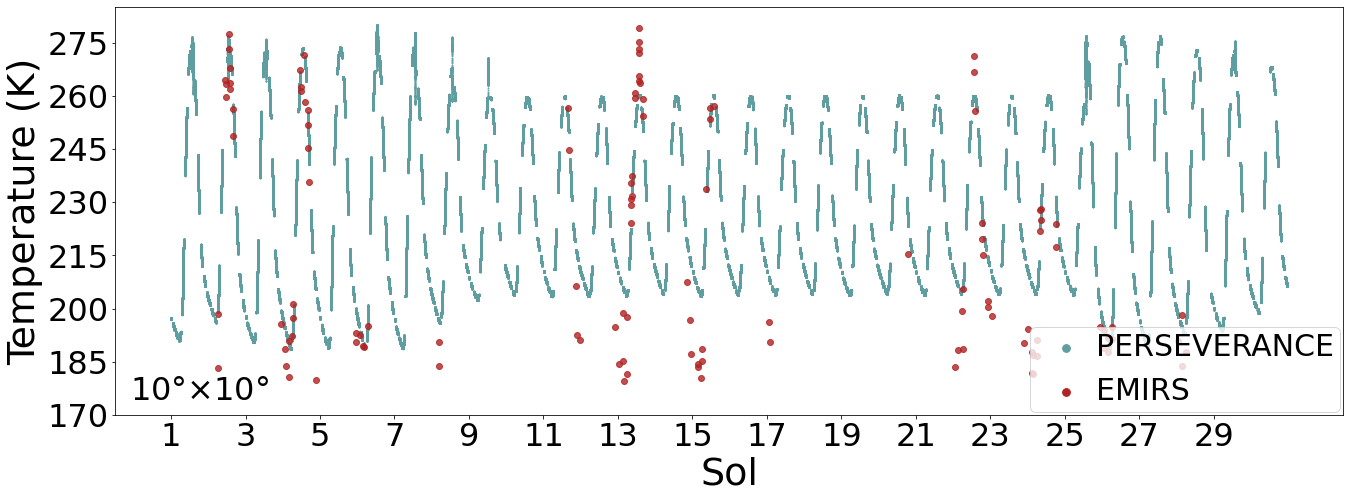}
 	\includegraphics[width=0.5\columnwidth]{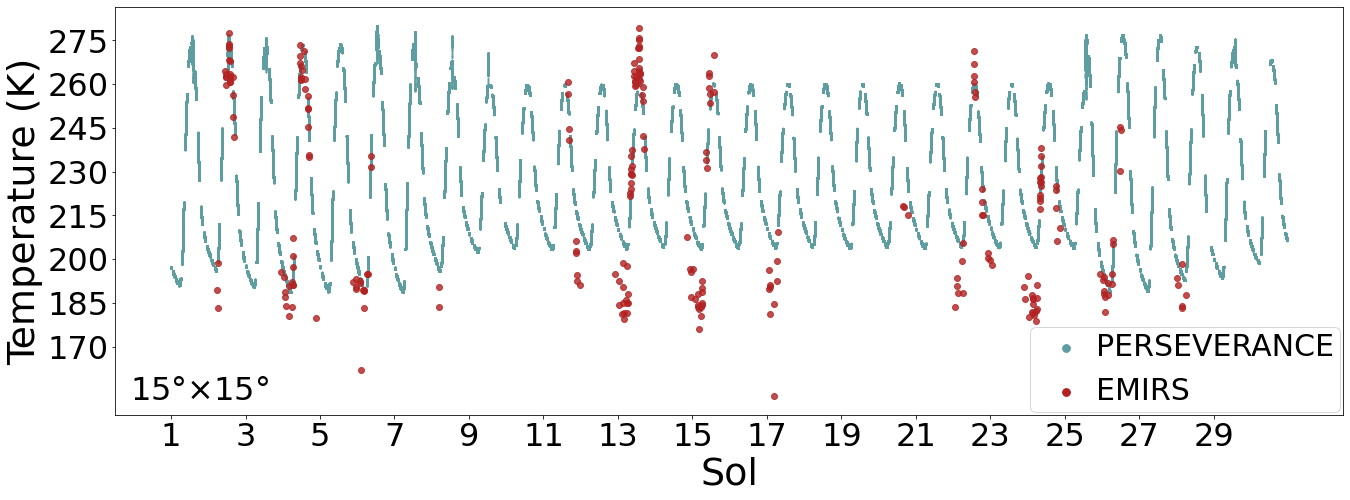}

     \caption{Comparison of hourly Perseverance and EMIRS surface temperature observations of various grid sizes --- $2\degree \times 2\degree$ ({\it top left}), $3\degree \times 3\degree$ ({\it top right}), $5\degree \times 5\degree$ ({\it middle left}), $7\degree \times 7\degree$ ({\it middle right}), $10\degree \times 10\degree$ ({\it bottom left}), $15\degree \times 15\degree$ ({\it  bottom right}) in July 2021.}
     \label{fig:comparison_grid4}
 \end{figure}

 \newpage
 \section{Diurnal coverage as a function of spatial grid size for selecting EMIRS observations about the location of the Curiosity and Perseverance rovers' landing sites.}
 %  \begin{figure}
 \makeatletter
\def\@captype{figure}
\makeatother
 \centering
 	\includegraphics[width=0.45\columnwidth,trim={0.8cm 0.5cm 1.8cm 1.5cm},clip]{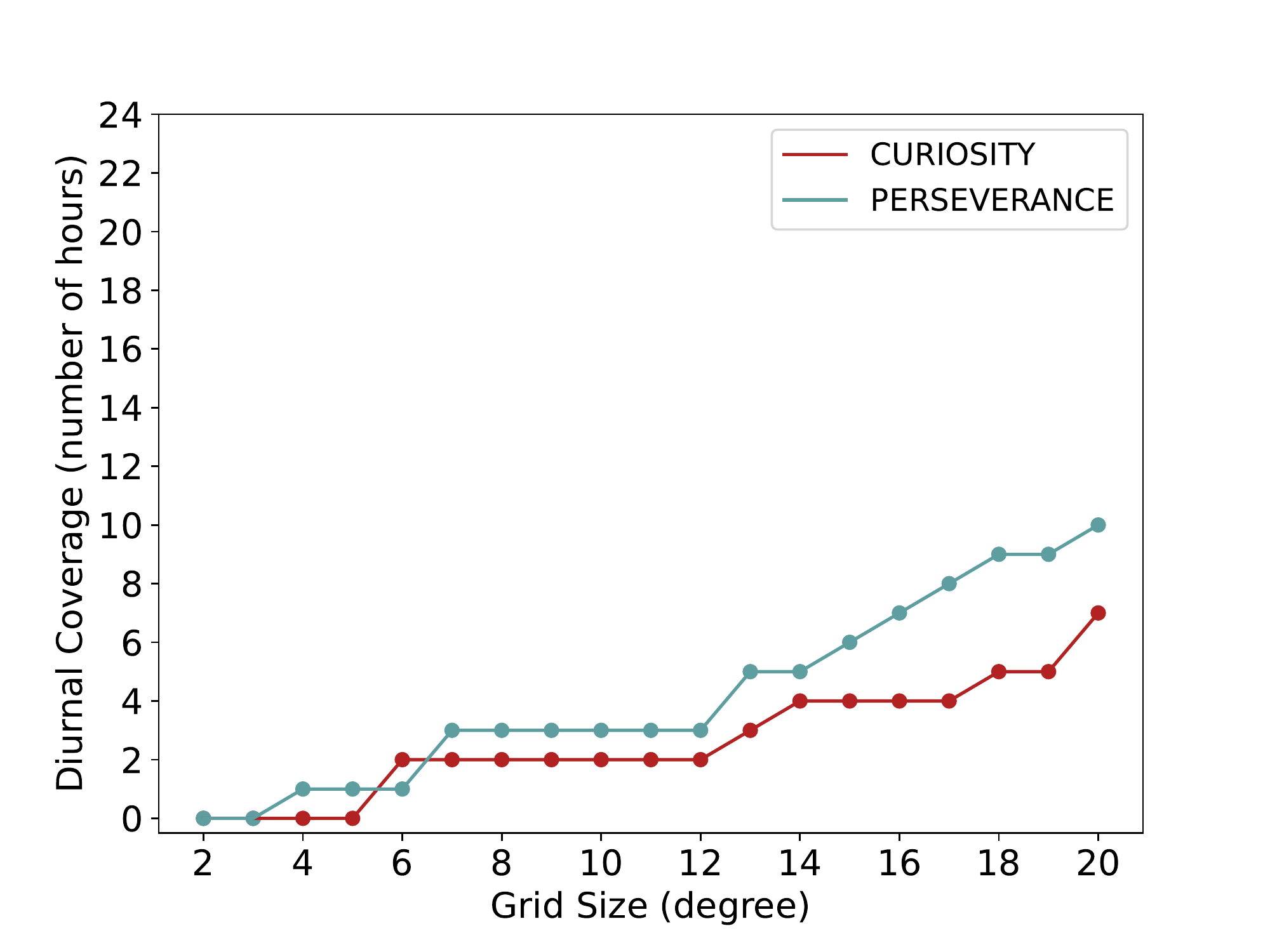}
 	 	\includegraphics[width=0.45\columnwidth,trim={0.8cm 0.5cm 1.8cm 1.5cm},clip]{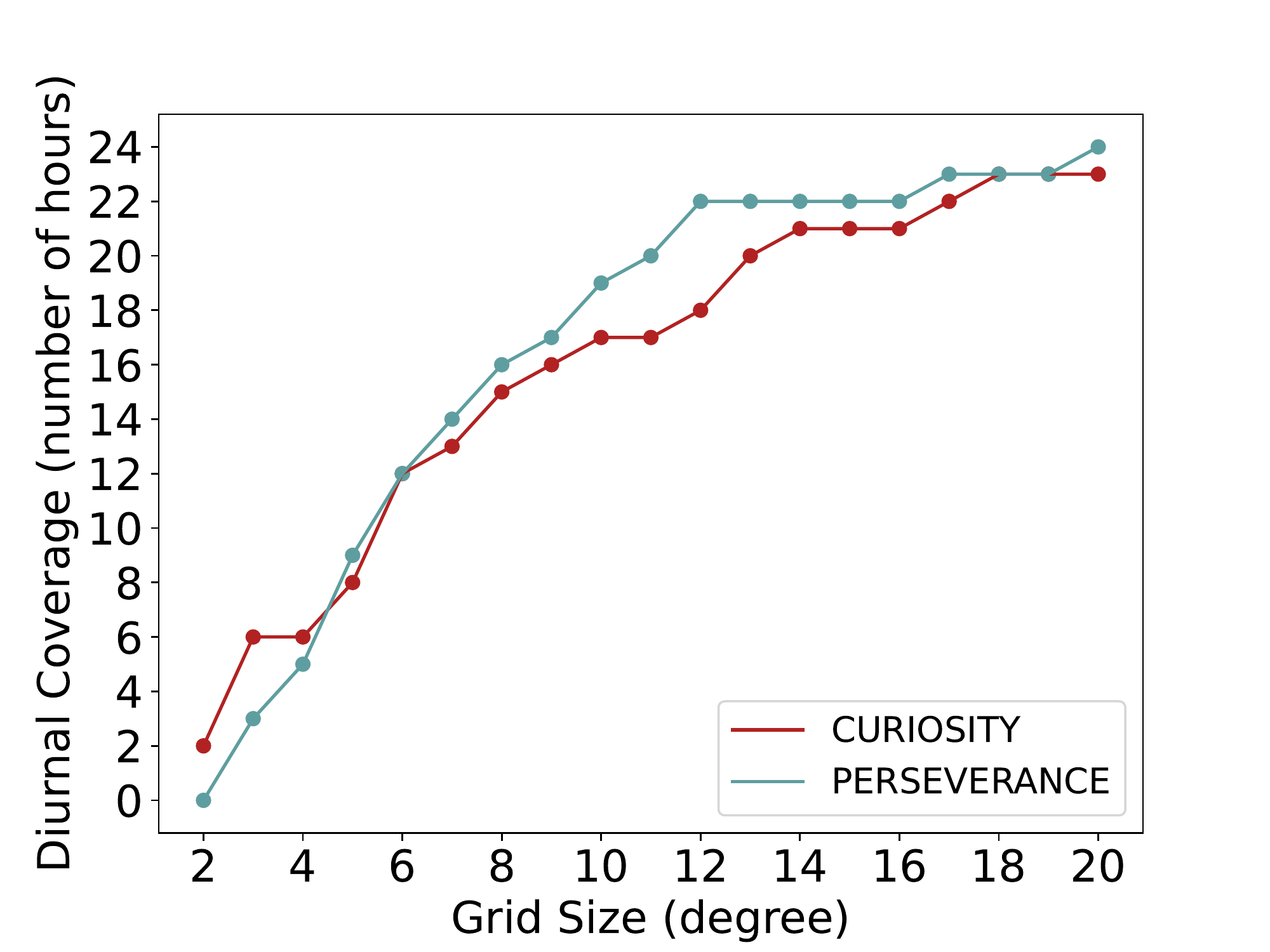}\\
 	\includegraphics[width=0.45\columnwidth,trim={0.8cm 0.5cm 1.8cm 1.5cm},clip]{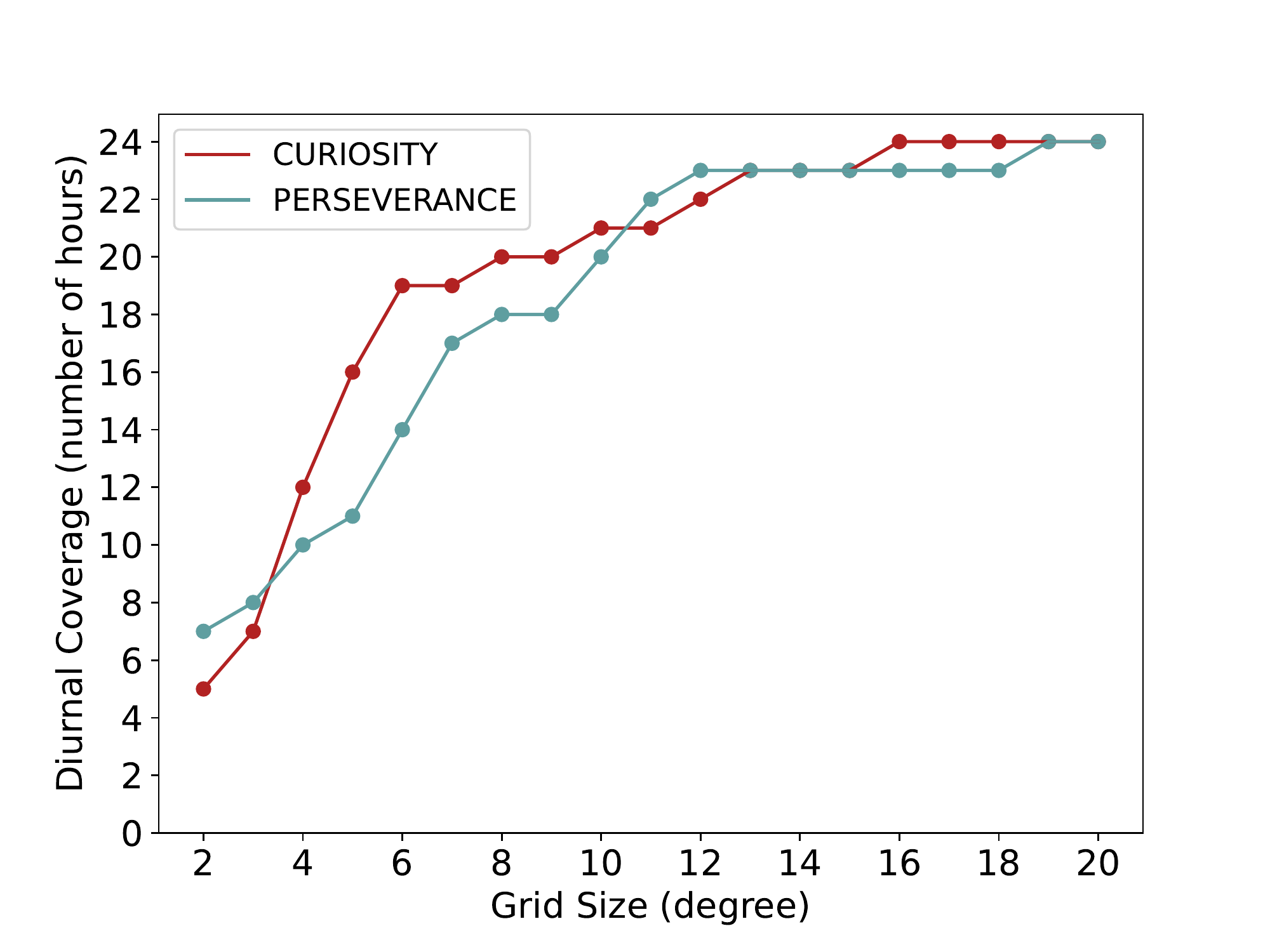}
 	\includegraphics[width=0.45\columnwidth,trim={0.8cm 0.5cm 1.8cm 1.5cm},clip]{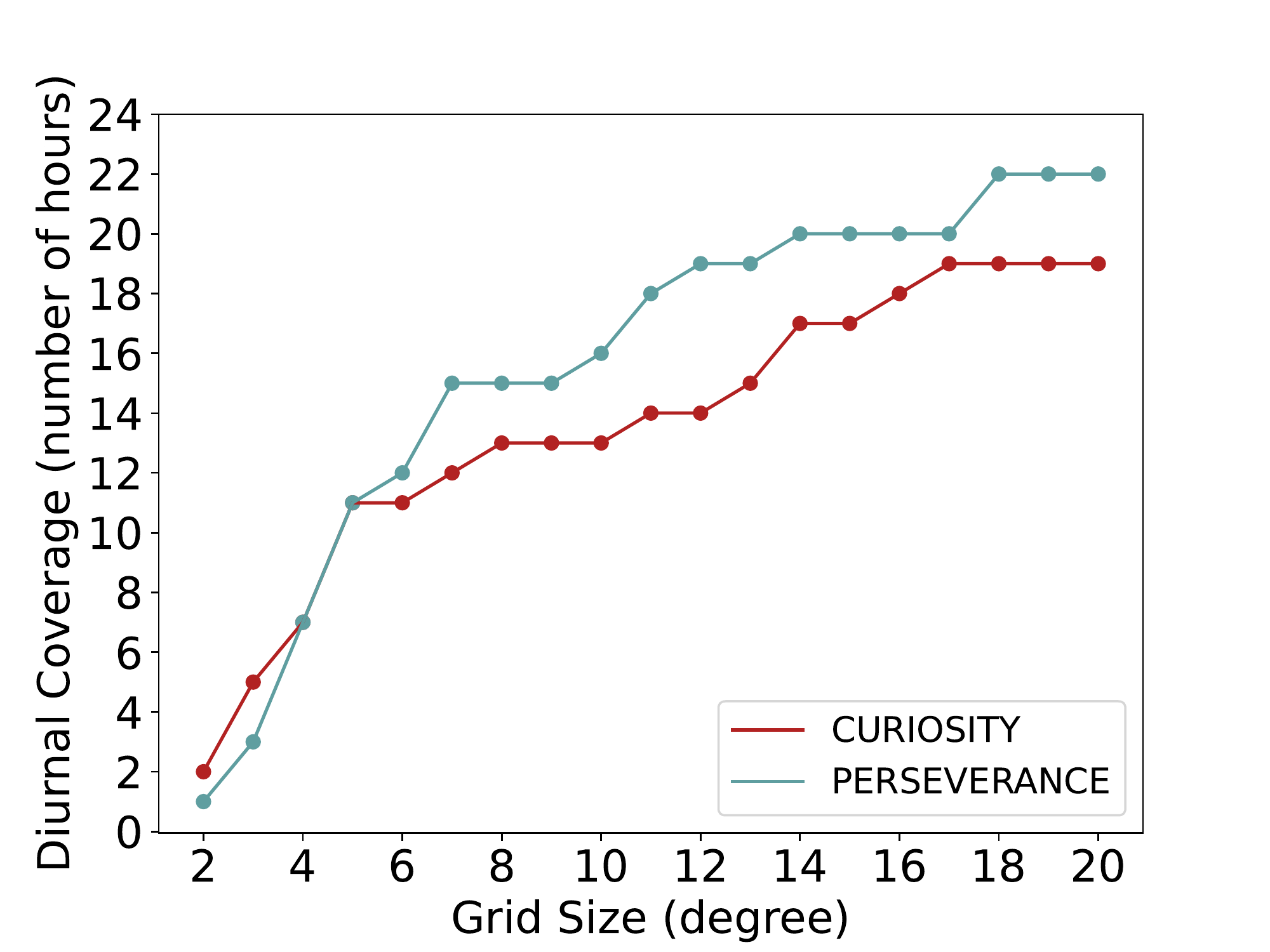}

     \caption{Diurnal coverage of EMIRS observations for various grid sizes at the location of Curiosity and Perseverance rovers --- May 2021 ({\it top left}), June 2021 ({\it top right}), July 2021 ({\it bottom left}), August 2021 ({\it bottom right}).}
     \label{fig:diurnal_coverage}
 %\end{figure}
 \makeatletter
\def\@captype{figure}
\makeatother
 \centering
 	\includegraphics[width=0.5\columnwidth]{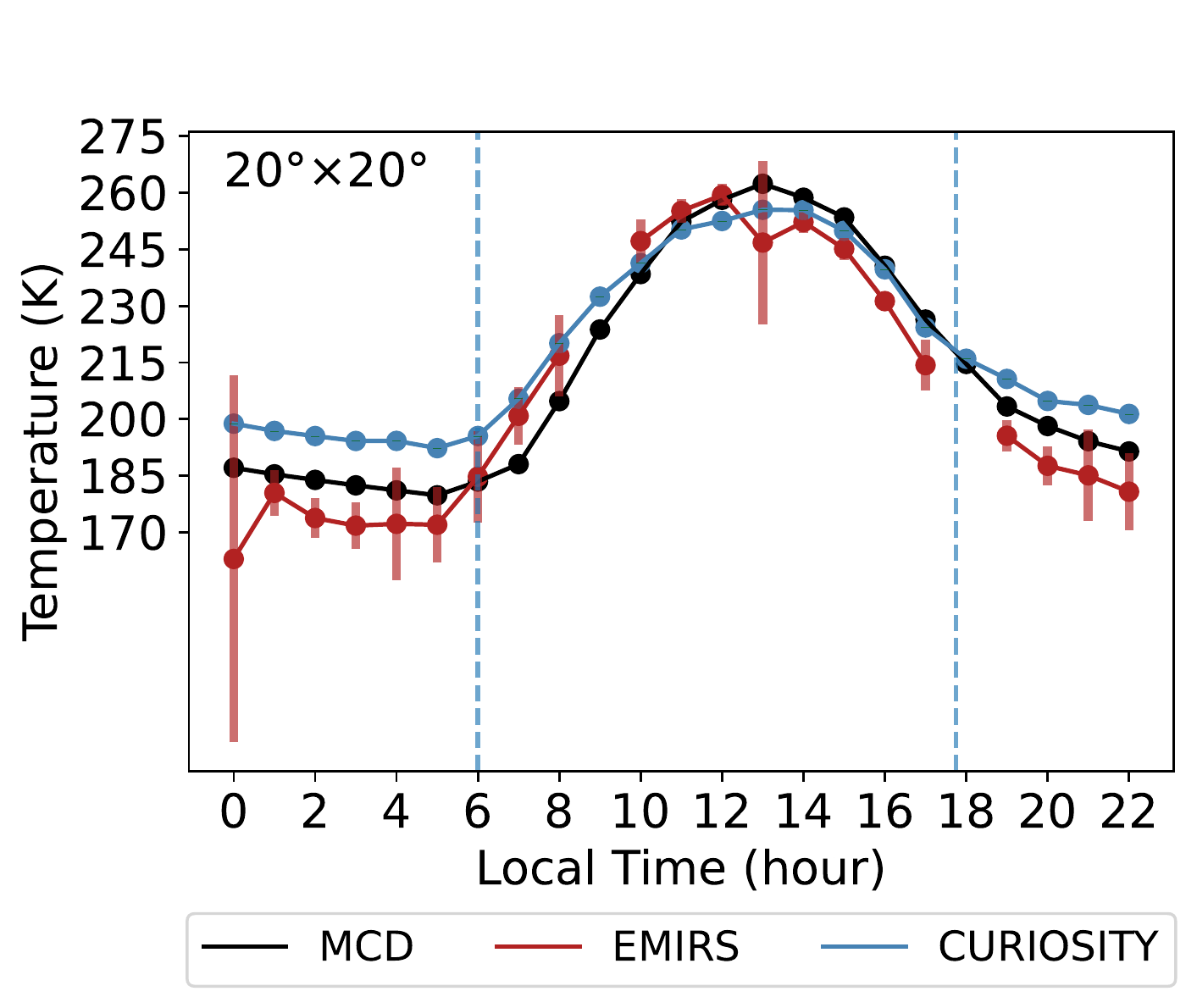}
     \caption{A comparison of average diurnal temperature variation over 30 sols measured by EMIRS and the MSL in starting L$_{s}$ = 52.6 (June 2021) with the model prediction from MCD. The MCD temperatures are obtained at the MSL rover site.}
     \label{fig:emirs_msl_june}
 \makeatletter
\def\@captype{figure}
\makeatother
 \centering
 	\includegraphics[width=0.5\columnwidth]{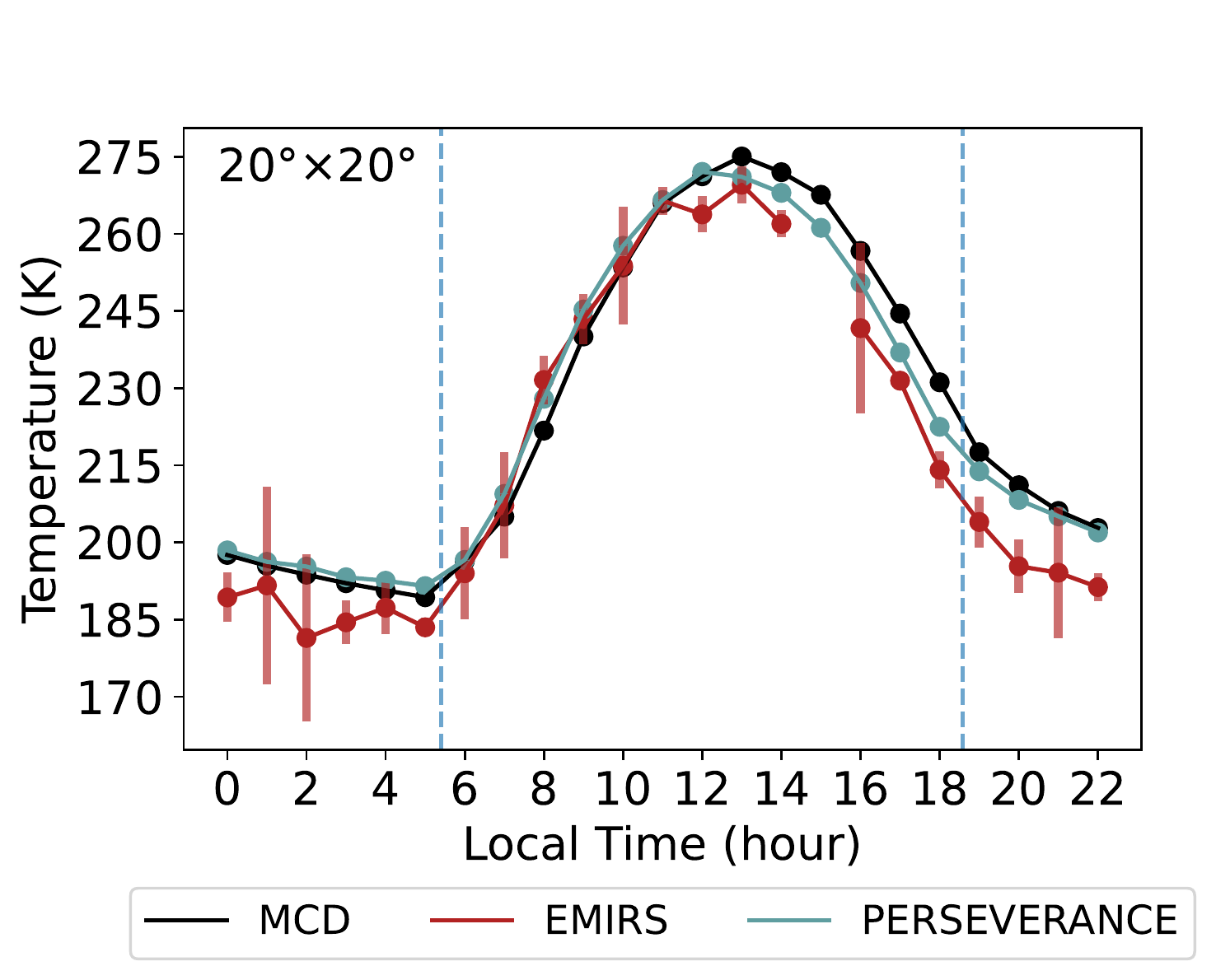}
     \caption{A comparison of average diurnal temperature variation over 30 sols measured by EMIRS and the Mars 2020 rover starting L$_{s}$ = 52.6 (June 2021) with the model prediction from MCD. The MCD temperatures are obtained at the Mars 2020 rover site.}
     \label{fig:emirs_percy_june}
 
 	\begin{figure}
 \centering
 	\includegraphics[width=0.45\columnwidth]{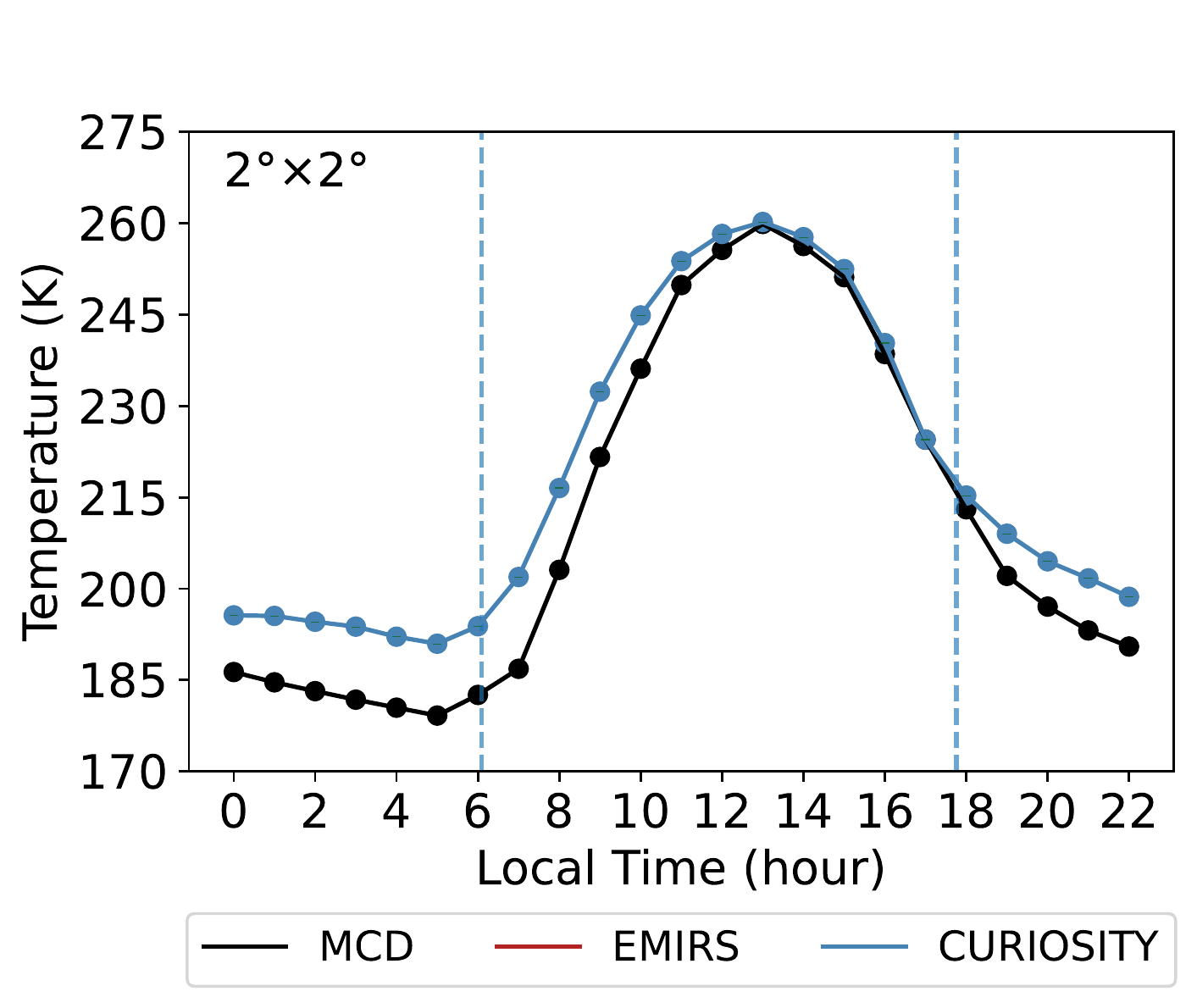}
  	\includegraphics[width=0.45\columnwidth]{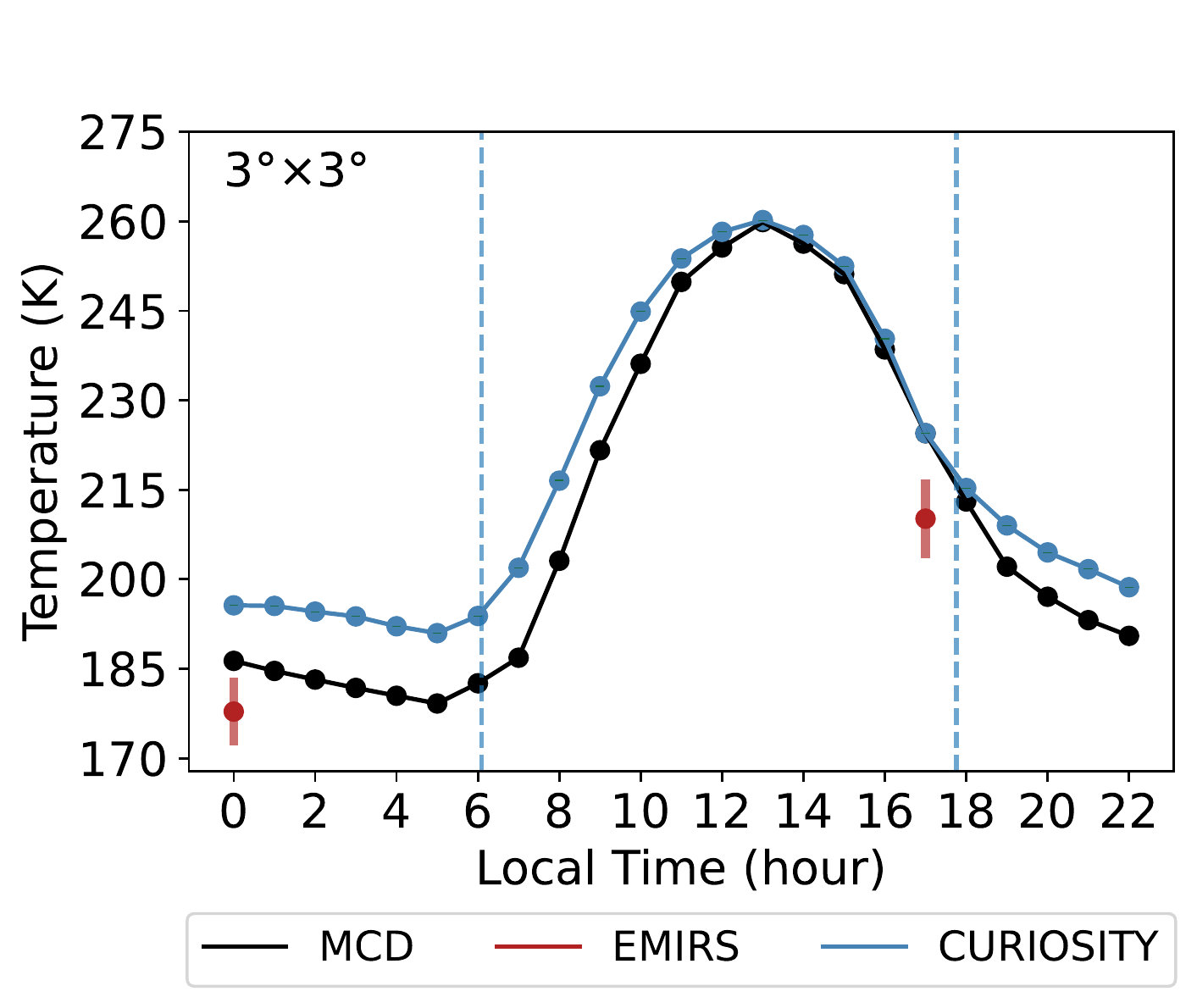}\\
  	 	\includegraphics[width=0.45\columnwidth]{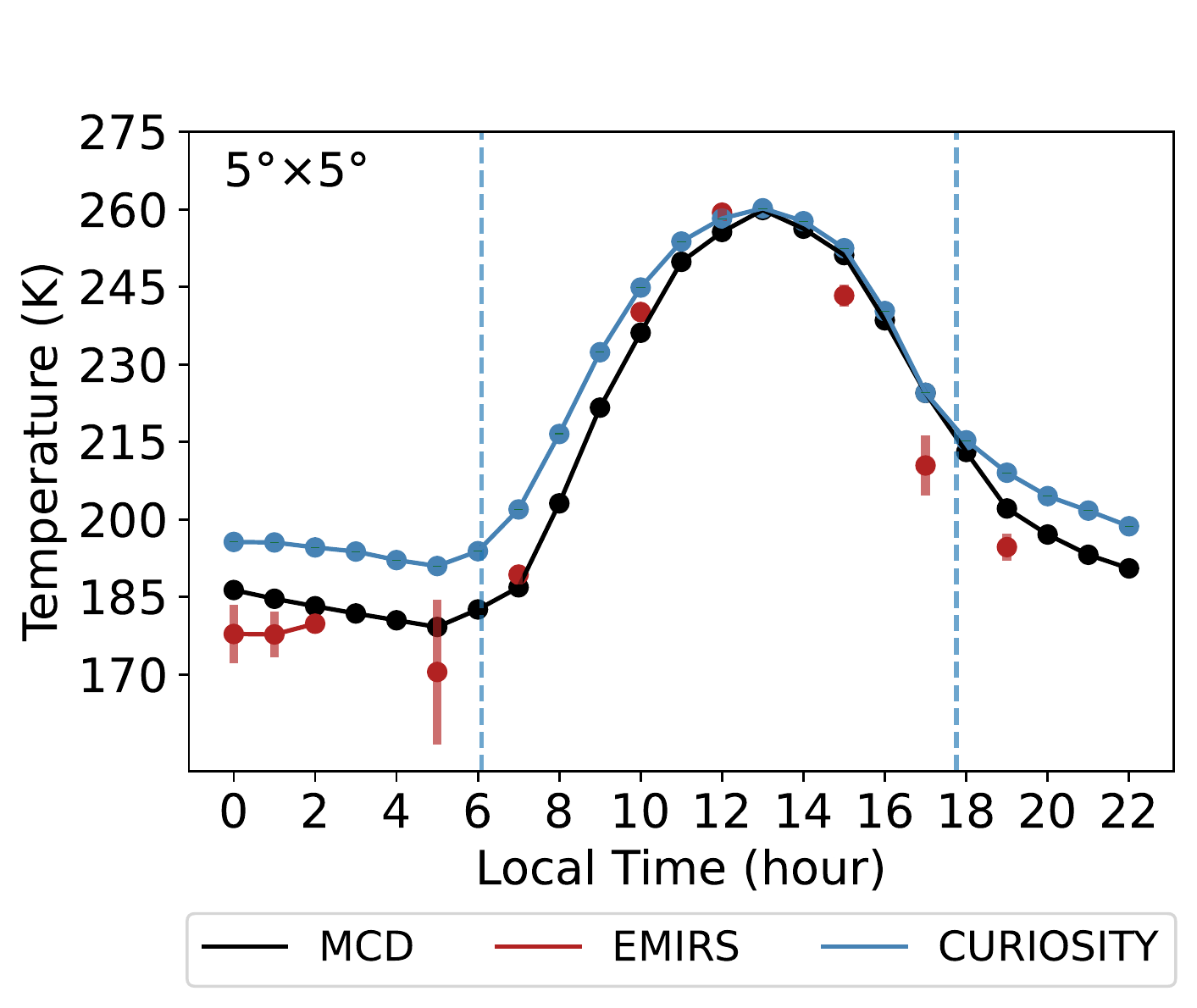}
	 	 	\includegraphics[width=0.45\columnwidth]{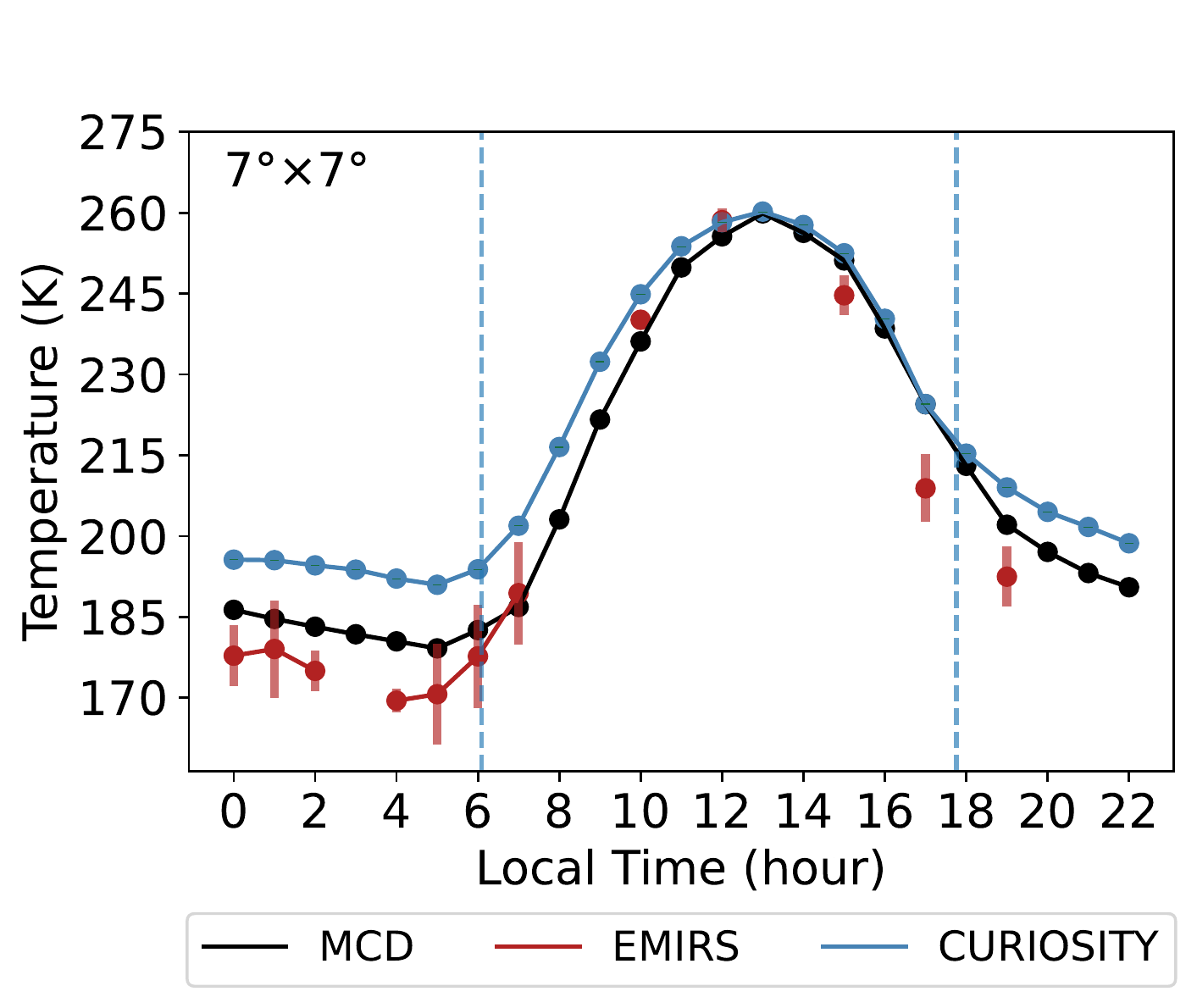}\\
  	 	 	 	\includegraphics[width=0.45\columnwidth]{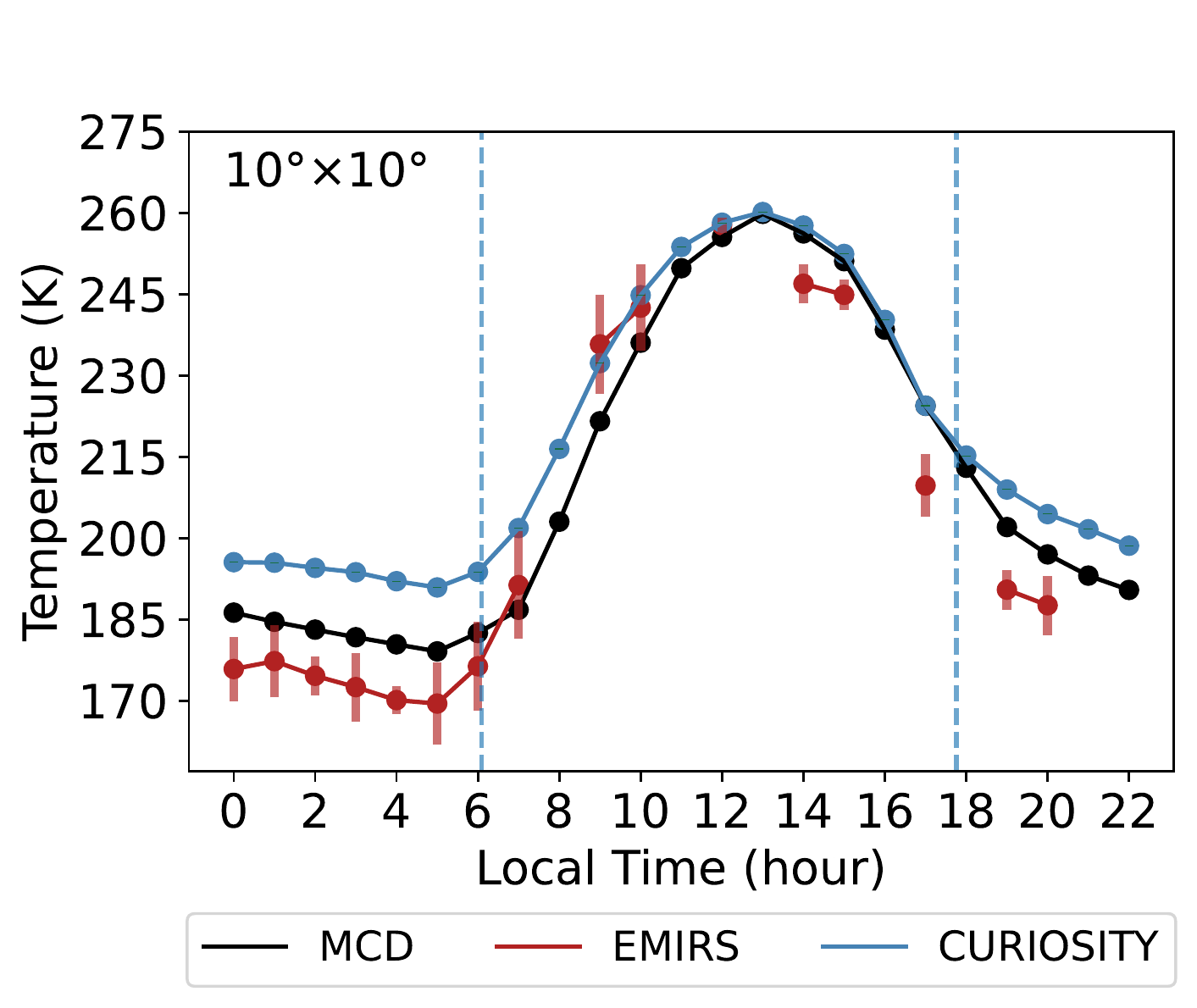}
  	 	 	 	 	\includegraphics[width=0.45\columnwidth]{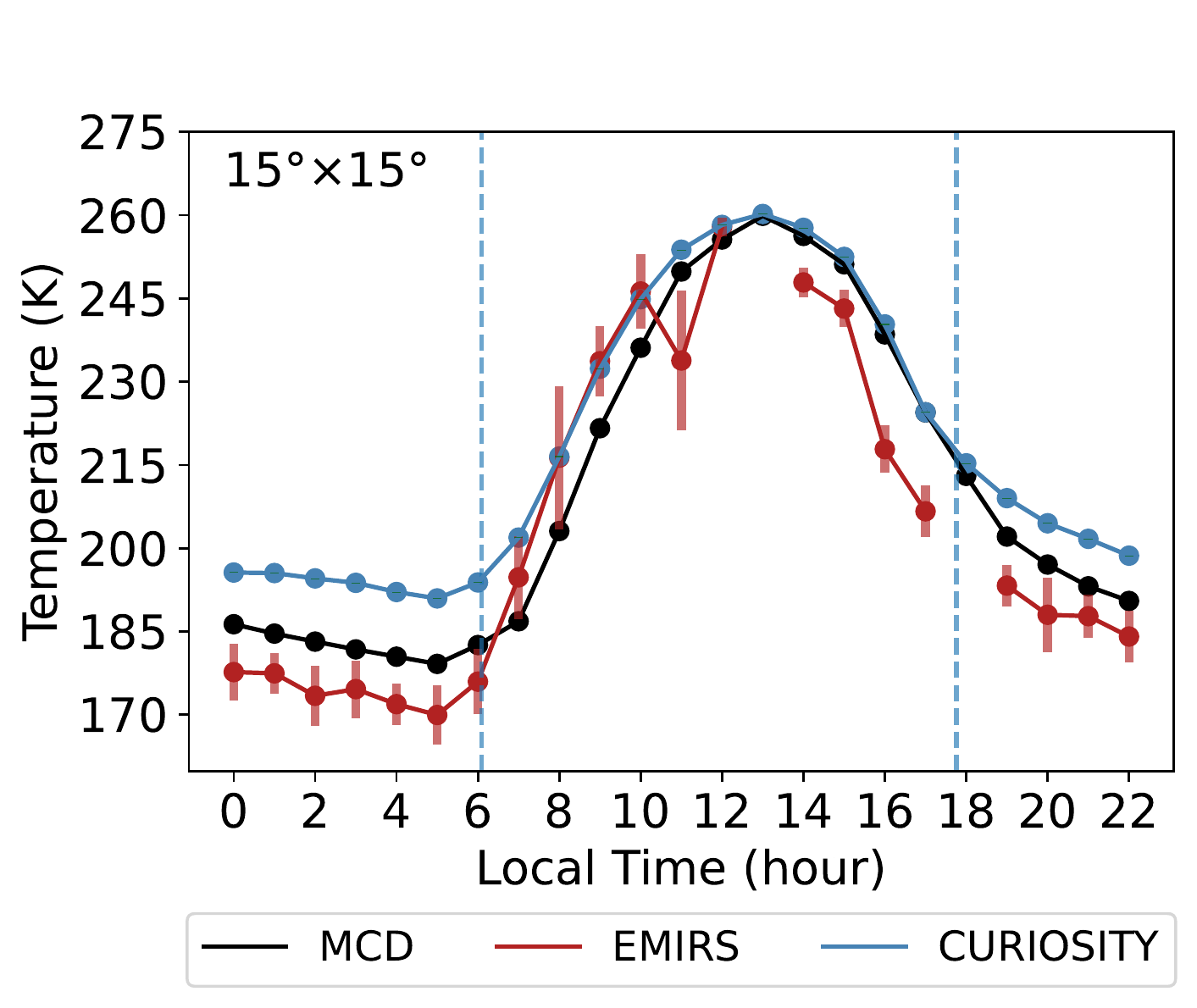}
     \caption{Average hourly surface temperature comparison as a function of EMIRS grid size at the MSL location --- $2\degree \times 2\degree$ ({\it top left}), $3\degree \times 3\degree$ ({\it top right}), $5\degree \times 5\degree$ ({\it middle left}), $7\degree \times 7\degree$ ({\it middle right}), $10\degree \times 10\degree$ ({\it bottom left}), $15\degree \times 15\degree$ ({\it  bottom right}).} \label{fig:comparison_grid1}
\end{figure}
 
     \begin{figure}
 \centering
 	\includegraphics[width=0.45\columnwidth]{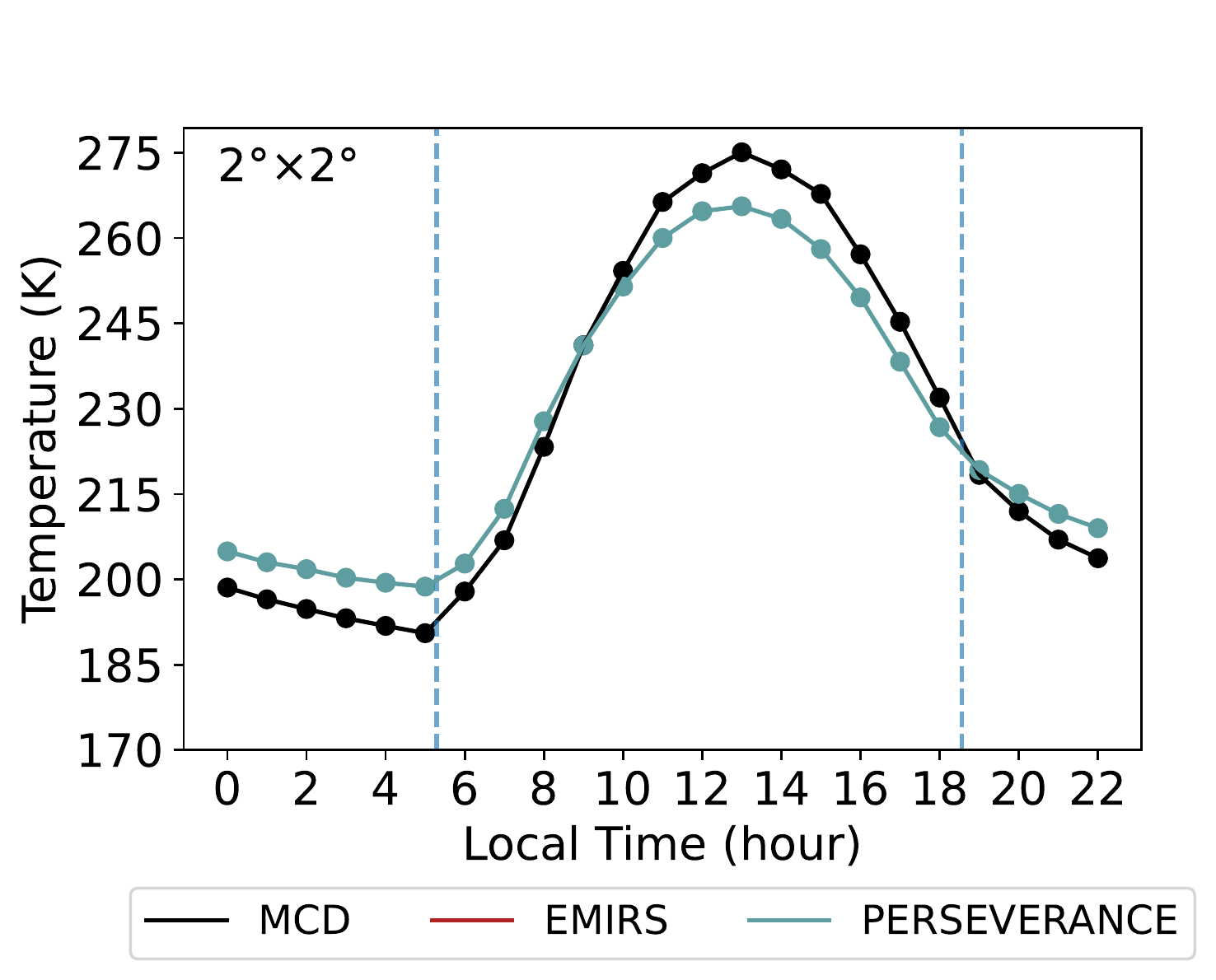}
  	\includegraphics[width=0.45\columnwidth]{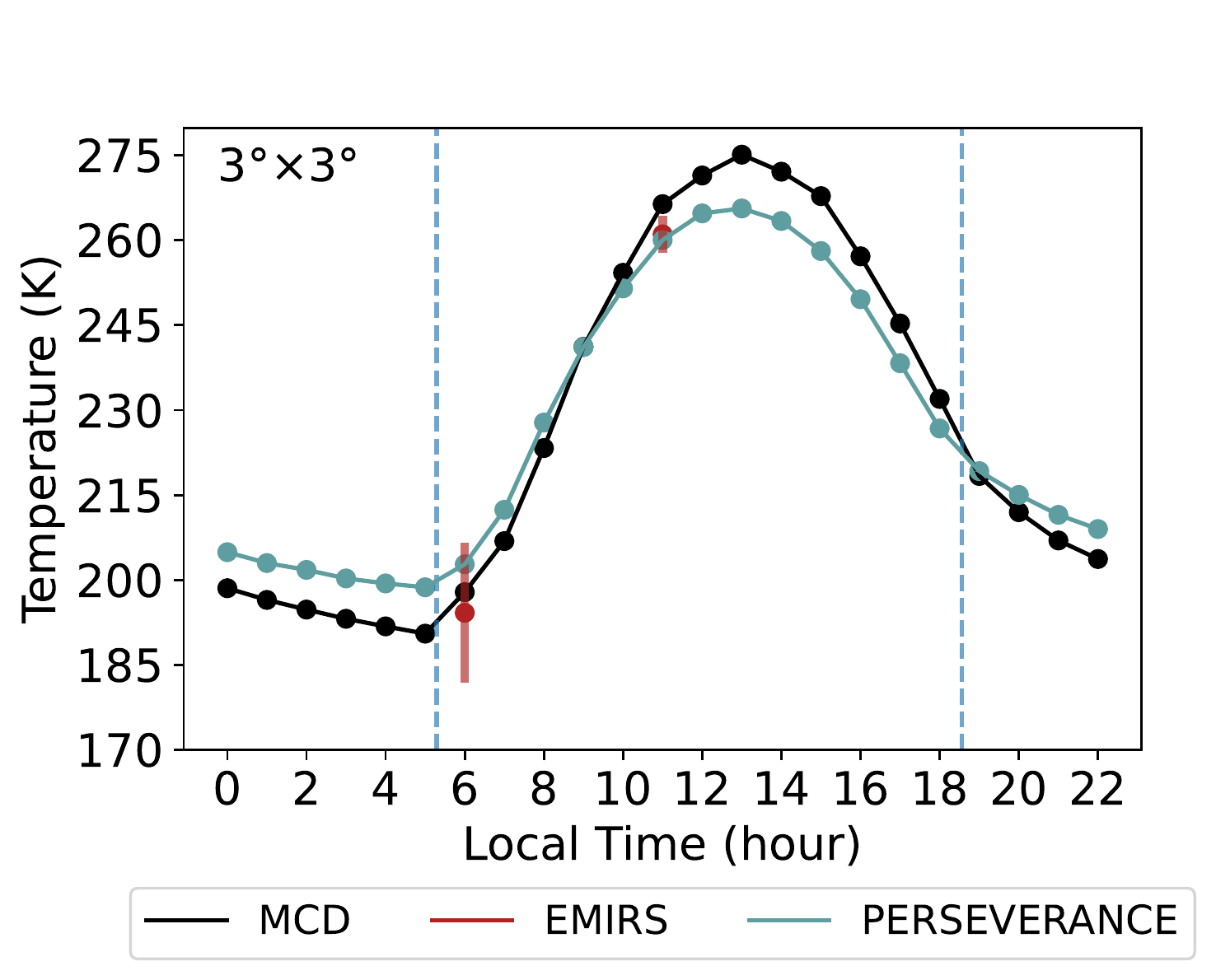}\\
  	 	\includegraphics[width=0.45\columnwidth]{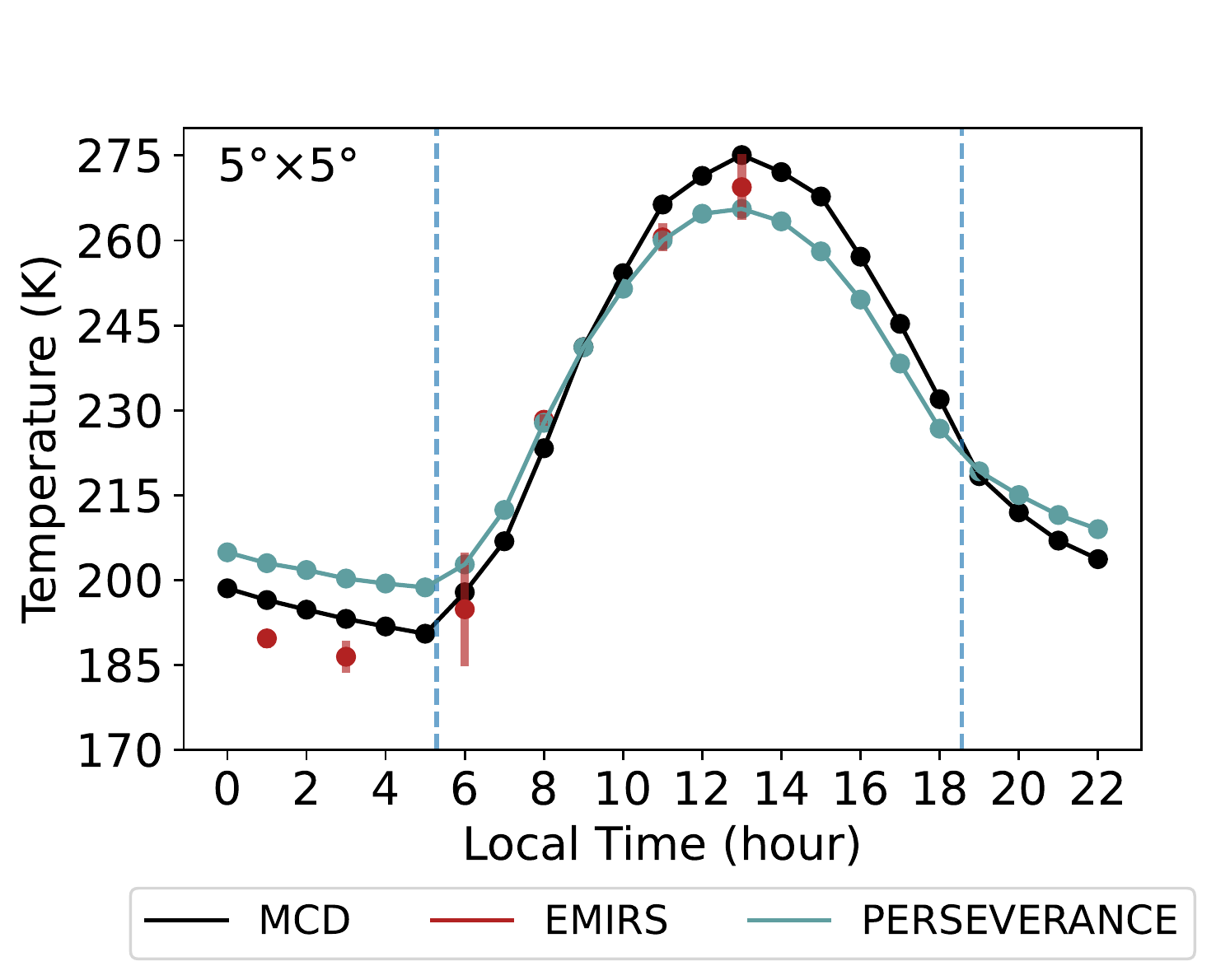}
  	 	\includegraphics[width=0.45\columnwidth]{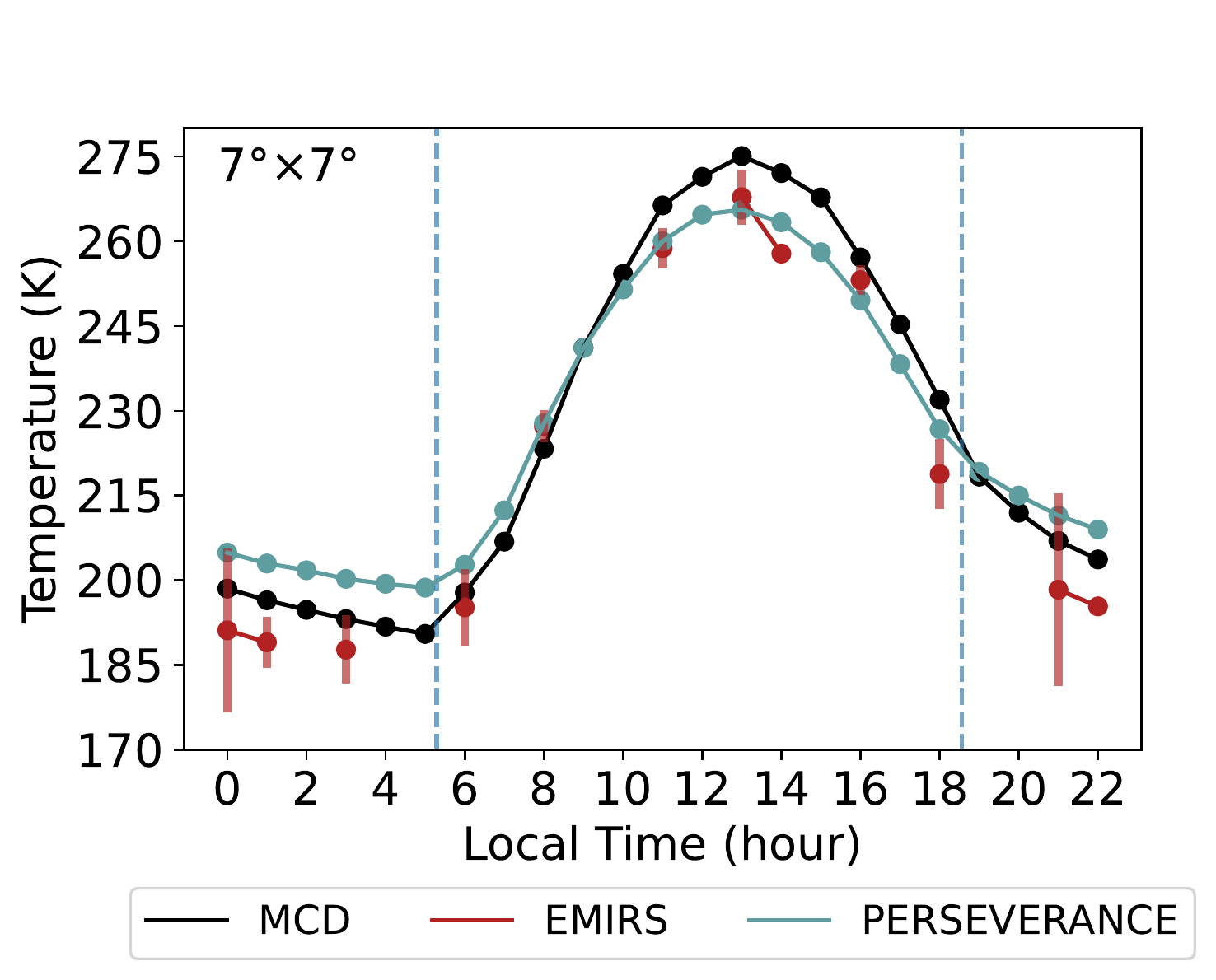}\\
  	 	 	 	\includegraphics[width=0.45\columnwidth]{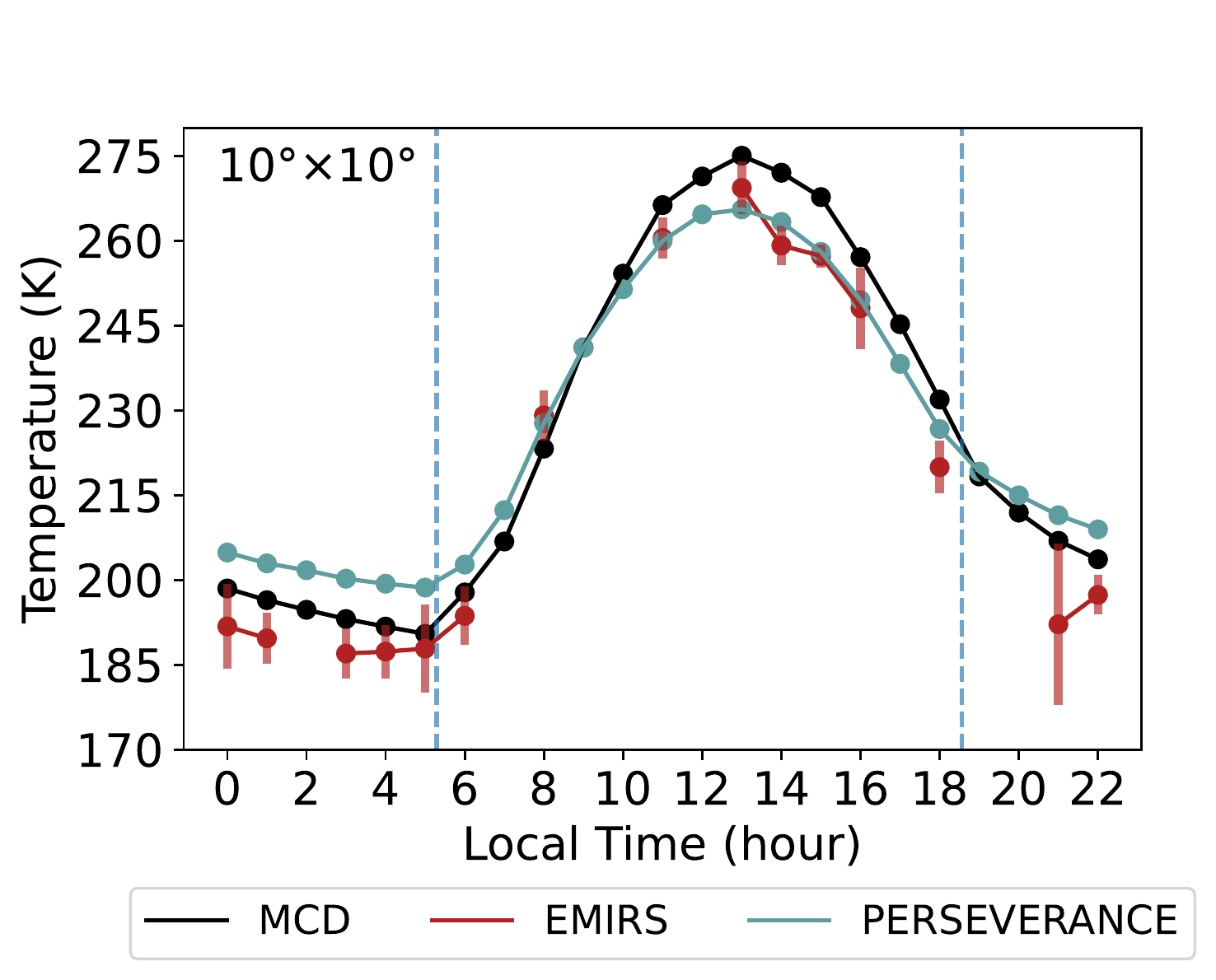}
  	 	 	 	 	\includegraphics[width=0.45\columnwidth]{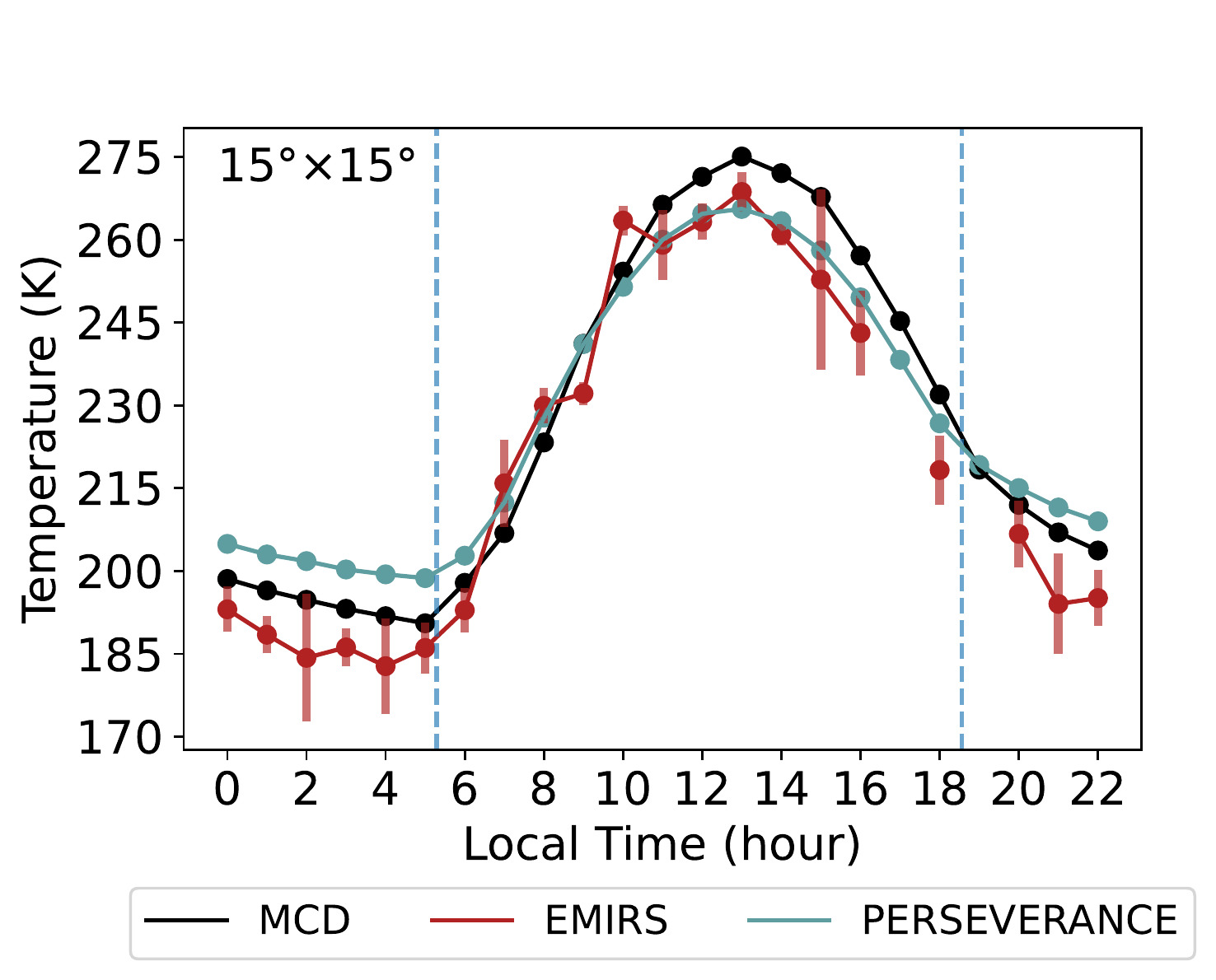}
     \caption{Average hourly surface temperature comparison as a function of EMIRS grid size at the Mars 2020 location --- $2\degree \times 2\degree$ ({\it top left}), $3\degree \times 3\degree$ ({\it top right}), $5\degree \times 5\degree$ ({\it middle left}), $7\degree \times 7\degree$ ({\it middle right}), $10\degree \times 10\degree$ ({\it bottom left}), $15\degree \times 15\degree$ ({\it  bottom right}).} 
     \label{fig:comparison_grid2}
 \end{figure}
 
 \newpage
 
 \section{Thermal Inertia Comparisons}
 \makeatletter
\def\@captype{figure}
\makeatother
 \centering
 	\includegraphics[width=0.50\columnwidth]{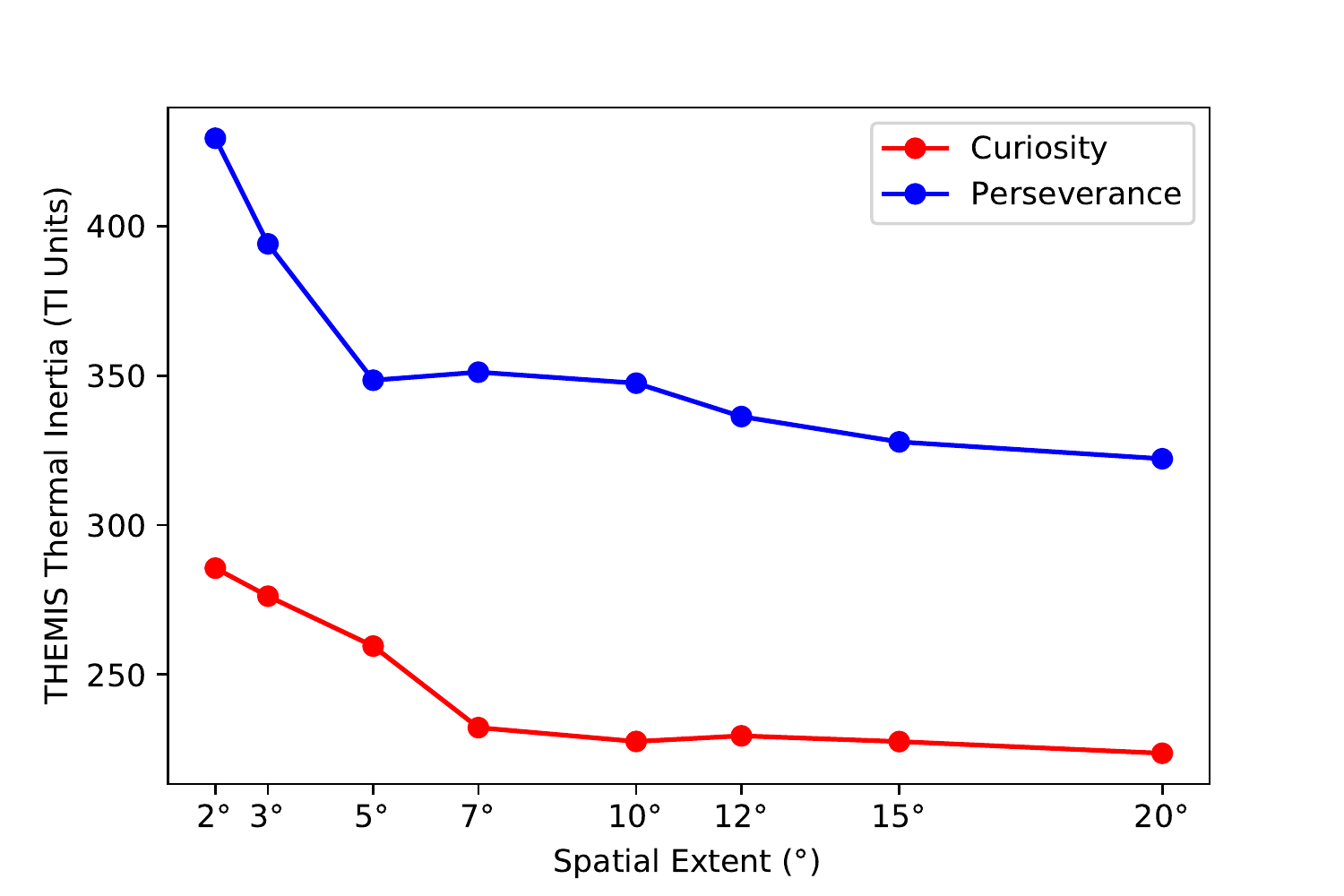}\\
 	\includegraphics[width=0.50\columnwidth]{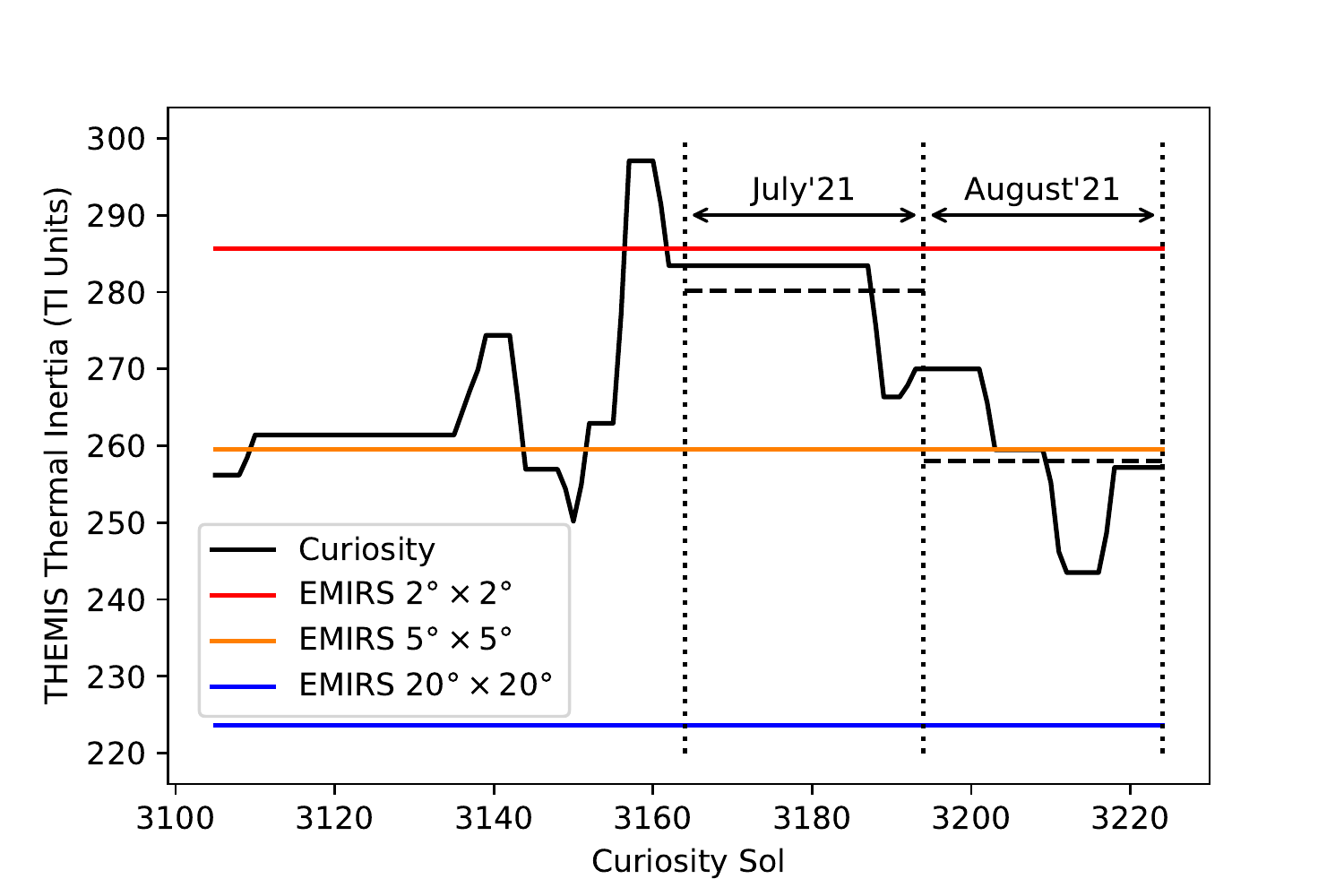}\\
 	\includegraphics[width=0.50\columnwidth]{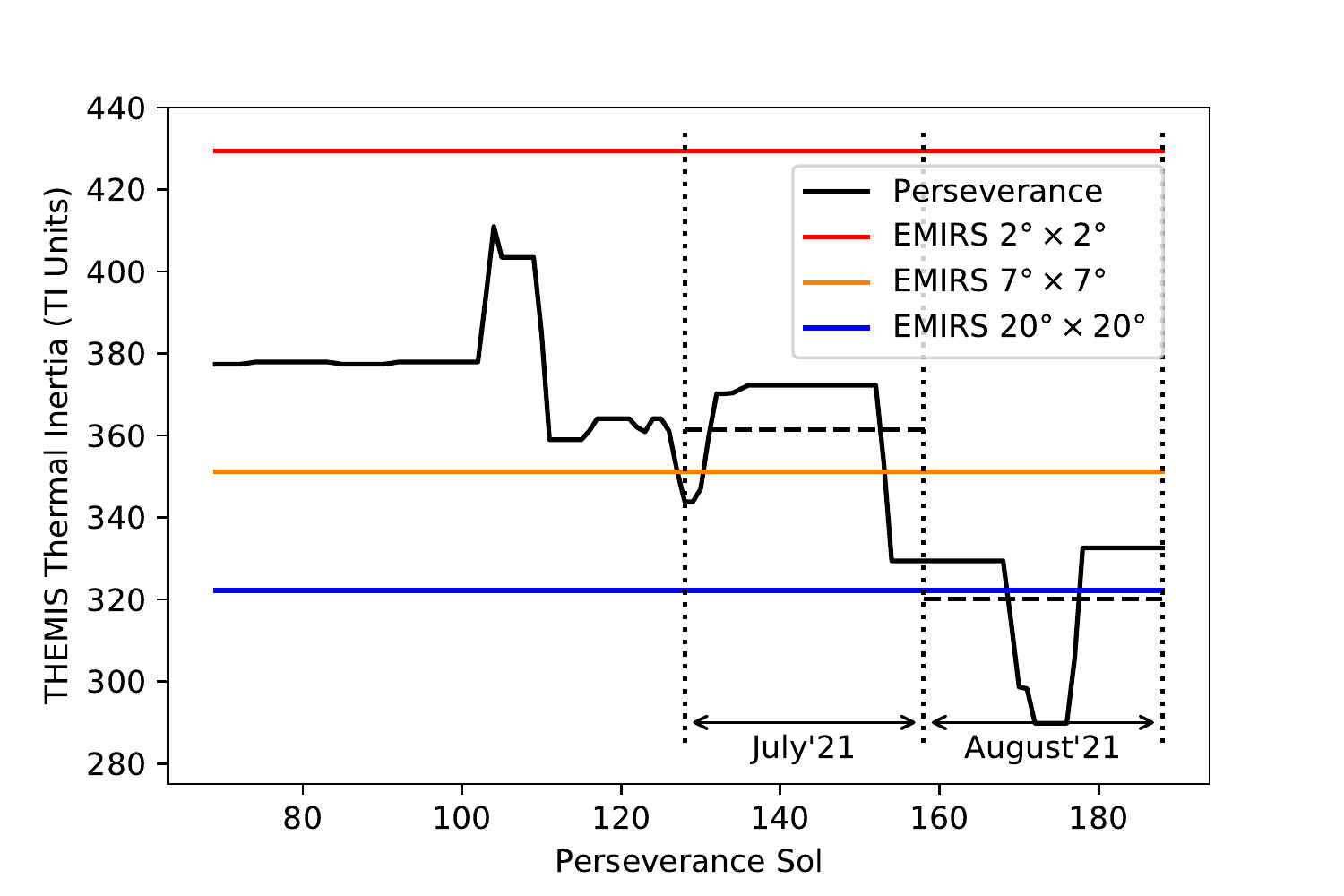}
     \caption{Comparison of Thermal Inertia (TI) obtained from THerMal Emission Imaging System (THEMIS) at the locations covered by the Curiosity and Perseverance rovers between May - August 2021 and the average TI obtained for the various EMIRS grid sizes centered at the landing sites of Curiosity and Perseverance. The dashed lines in the bottom two panels show the average TI values for the respective months for the locations covered by the respective rovers.}
     \label{fig:thermal_inertia}


 %%%%%%%%%%%%%%%%%%%%%%%%%%%%%%%%%%%%%%%%%%%%%%%%%%

 % Don't change these lines
\bsp	% typesetting comment
\label{lastpage}
\end{document}